\documentclass[letterpaper,twocolumn,10pt]{article}
\usepackage{usenix}

\usepackage{tikz}
\usepackage{amsmath}
\usepackage{amssymb}
\usepackage{tcolorbox}
\usepackage{graphicx,subfigure}
\usepackage{color, colortbl}
\usepackage{subfigure}
\usepackage{adjustbox}
\usepackage{multirow}
\usepackage{tabularx}
\usepackage{makecell}
\usepackage{enumerate}
\usepackage{xurl} 
\usepackage{xcolor,pifont}
\usepackage[linesnumbered,ruled,lined]{algorithm2e}
\usepackage{wasysym}
\usepackage{arydshln}
\usepackage[euler]{textgreek}

\usepackage{caption} 
\usepackage[toc,page]{appendix}

\newcolumntype{a}{>{\columncolor{Gray}}c}
\definecolor{Gray}{gray}{0.85}

\SetKwInput{KwInput}{Input}                
\SetKwInput{KwOutput}{Output}              

\definecolor{LightGray}{gray}{0.9}

\def\BibTeX{{\rm B\kern-.05em{\sc i\kern-.025em b}\kern-.08em
    T\kern-.1667em\lower.7ex\hbox{E}\kern-.125emX}}   

\newcommand*\colourcheck[1]{%
  \expandafter\newcommand\csname #1check\endcsname{\textcolor{#1}{\ding{52}}}%
}
\colourcheck{green}

\newcommand*\colorx[1]{%
  \expandafter\newcommand\csname #1x\endcsname{\textcolor{#1}{\ding{55}}}%
}
\colorx{red}

\definecolor{Gray}{gray}{0.9}

\usepackage{filecontents}

\begin{document}

\date{}
\title{Demoting Security via Exploitation of Cache Demote Operation in Intel’s Latest ISA Extension}


\author{
{\rm Taehun Kim}\\
Korea University
\and
{\rm Hyerean Jang}\\
Korea University
\and
{\rm Youngjoo Shin}\\
Korea University}

\maketitle

\begin{abstract}
ISA extensions are increasingly adopted to boost the performance of specialized workloads without requiring an entire architectural redesign.
However, these enhancements can inadvertently expose new attack surfaces in the microarchitecture.
In this paper, we investigate Intel’s recently introduced \texttt{cldemote} extension, which promotes efficient data sharing by transferring cache lines from upper-level caches to the Last Level Cache (LLC).
Despite its performance benefits, we uncover critical properties—unprivileged access, inter-cache state transition, and fault suppression—that render \texttt{cldemote} exploitable for microarchitectural attacks.
We propose two new attack primitives, Flush+Demote and Demote+Time, built on our analysis.
Flush+Demote constructs a covert channel with a bandwidth of 2.84~Mbps and a bit error rate of 0.018\%, while Demote+Time derandomizes the kernel base address in 2.49~ms on Linux.
Furthermore, we show that leveraging \texttt{cldemote} accelerates eviction set construction in non-inclusive LLC designs by obviating the need for helper threads or extensive cache conflicts, thereby reducing construction time by 36\% yet retaining comparable success rates.
Finally, we examine how ISA extensions contribute to broader microarchitectural attacks, identifying five key exploitable characteristics and categorizing four distinct attack types.
We also discuss potential countermeasures, highlighting the far-reaching security implications of emerging ISA extensions.
\end{abstract}

\section{Introduction}
\label{sec:introduction}

Modern processors are continually evolving to support a wider range of workloads more efficiently.
A key driver of this evolution is the introduction of Instruction Set Architecture (ISA) extensions, which expand a processor’s capabilities without requiring a complete redesign.
For example, x86 has integrated Streaming SIMD Extensions (SSE) and Advanced Vector Extensions (AVX) into its architecture, enabling efficient handling of large-scale scientific tasks.

While ISA extensions offer significant benefits, they can also create unintended interactions between software and hardware layers.
This arises because such extensions often interact with microarchitectural components—such as the cache~\cite{yarom2014flush+, liu2015last, gruss2016flush+} and execution units~\cite{aldaya2019port, bhattacharyya2019smotherspectre}—which are known to be susceptible to side-channel attacks.
These extensions can open new attack surfaces to potential adversaries.

In this paper, we analyze Intel’s new \texttt{cldemote} extension, which is featured in its latest processors, including server-grade Xeon and the low-power Atom.
Intel introduced this extension to improve performance in multi-threaded workloads that frequently share data~\cite{Intel2024AORM, Intel2024IASEFF, Intel2024Lab}.
The primary function of \texttt{cldemote} is to move a cache line from an upper-level cache  to a lower-level cache, thereby increasing the likelihood of Last-Level Cache (LLC) hits rather than remote cache hits and enhancing the efficiency of shared data access.

Despite its performance benefits, the security implications of \texttt{cldemote} are largely unexplored.
In line with this, we conducted an in-depth analysis to identify potential vulnerabilities that could be exploited in microarchitectural attacks.
Our findings indicate that \texttt{cldemote} can be invoked at unprivileged privilege levels, induces microarchitectural state transitions between private caches and the shared LLC, and includes a fault suppression feature (vid. Section~\ref{sec:key_property}).

Based on these observations, we investigate the security implication and present two case studies:
\begin{enumerate}
\item \textbf{Side-channel attacks} (vid. Section~\ref{sec:case_study1}).
We show that \texttt{cldemote} can be leveraged for side-channel attacks. 
Specifically, we propose two novel attack primitives by using the extension.
\item \textbf{Eviction set construction} (vid. Section~\ref{sec:case_study2}).
We further show that using \texttt{cldemote} can streamline eviction set construction in non-inclusive LLC designs, overcoming limitations in prior methods~\cite{yan2019attack, kim2022dprime+, zhao2024last, purnal2021prime+}.
\end{enumerate}

For side-channel attacks, we build two attack primitives by using the \texttt{cldemote} extension.
First, Flush+Demote allows an unprivileged attacker to infer the cache state by measuring timing differences in the execution latency of \texttt{cldemote}.
Second, Demote+Time enables the attacker to distinguish valid from invalid kernel addresses by inferring the TLB state.
We demonstrate the effectiveness of these primitives through two practical attack scenarios.
Specifically, Flush+Demote can construct a covert channel with a bandwidth of 2.84~Mbps and a bit error rate of 0.018\%. 
We also show that Demote+Time derandomizes a kernel base address in 2.49~ms on Linux.

Regarding eviction set construction, it is challenging to devise the method in non-inclusive LLC, where only cache lines in private caches are allocated for memory accesses.
To force moving cache lines from upper to lower-level cache, previous approaches require huge number of memory operations~\cite{yan2019attack} or additional cores~\cite{purnal2021prime+, zhao2024last, yan2019attack}.
We leverage the property of \texttt{cldemote} that directly transfers a cache line to the LLC to build our efficient eviction set constructing technique.
Our technique overcomes the limitations of the previous eviction set constructing approaches  by eliminating the need for multiple memory operations or additional cores.

To implement the eviction set construction technique, detailed knowledge of the microarchitecture of LLC, such as set associativity and directory structure, is necessary, which is not publicly available for the Intel's new architecture.
Hence, we reverse-engineered the hidden microarchitectural details on Sapphire Rapids processors.
Our novel eviction set construction technique for non-inclusive LLC, dubbed Access+Demote, simplifies the construction process and significantly accelerates the overall build time.
Our proof-of-concept implementation shows that replacing older techniques with \texttt{cldemote} decreases construction time by 36\% while maintaining comparable success rates, as measured against the algorithms of Vila et al.~\cite{vila2019theory} and Zhao et al.~\cite{zhao2024last}.
We also discuss potential mitigations against the proposed attacks.

In addition to case studies, we present a comprehensive analysis of microarchitectural side-channel attacks~\cite{gruss2016flush+, yarom2014flush+,gruss2016prefetch, lipp2022prefetch, guo2022adversarial,weber2021osiris, choi2023avx,kim2023avx, jang2016breaking, kim2022dprime+, disselkoen2017prime+} that exploit ISA extensions (vid. Section~\ref{sec:ISA extension driven}). 
We analyze how these extensions interact with microarchitectural components.
From the analysis, we identify five key exploitable characteristics:
(1)~unprivileged access,
(2)~inter-cache state transition,
(3)~cache-to-memory state transition,
(4)~cache state transition detection, and
(5)~fault suppression.
Combining these characteristics yields four distinct attack categories: cache attacks, noise-free cache attacks, faultless KASLR breaking, and fast eviction set construction.
We refer to such attacks collectively as \emph{ISA extensions-driven attacks}.
This broader lens demonstrates that \texttt{cldemote} is far from an isolated case; similar characteristics in other ISA extensions can lead to equally potent security risks.

\smallskip
\noindent\textbf{Contributions.}
Our main contributions are as follows:
\vspace{-0.1cm}
\begin{itemize}
\setlength{\itemsep}{-0.3em}
\item We analyze the architectural and microarchitectural properties of Intel's new \texttt{cldemote} extension.
\item We present two practical attacks: a covert channel attack using Flush+Demote and a KASLR-breaking attack on Linux using Demote+Time.
\item We successfully reverse-engineered the LLC structure on Intel Sapphire Rapids processors, providing the foundation for our eviction set construction technique.
\item We present an efficient eviction set construction technique that overcomes the limitations of existing methods for a non-inclusive LLC, reducing construction time by 36\% while maintaining comparable success rates.
\item We identify five key characteristics of various ISA extensions that make them exploitable.
We then categorize four distinct attack types based on these characteristics.
\end{itemize}
\vspace{-0.1cm}


\section{Background}
\label{sec:background}

\subsection{Cache organization}
Modern processors employ a multi-level cache hierarchy to compensate for the slow access latency from the main memory.
The cache hierarchy typically consists of 1) Private caches (L1 and L2) within each core
and 2) a shared Last Level Cache (LLC) in the processor's uncore, divided into LLC slices, each corresponding to a core.
In this hierarchy, data stored closer to the core is accessed with lower latency.

A key attribute in the cache architecture is inclusiveness.
Intel initially adopted an inclusive LLC design, where any cache line present in a private cache is also stored in the LLC.
However, as core counts increased, this led to inefficient LLC utilization due to redundant data from private caches.
To address this, Intel Xeon processors, starting with Skylake-SP, transitioned to a non-inclusive LLC design, optimizing space and reducing coherence traffic.

Since the non-inclusive LLC architecture, Intel introduced a directory structure in its coherence protocol to maintain data consistency~\cite{gupta1992reducing}.
This directory tracks cache lines in private caches, making it inclusive for private caches, while the LLC manages only its own lines.
The directory, partitioned into directory slices per core, is organized as a set-associative cache and uses the same address mapping hash function as the LLC to locate slices and sets~\cite{yan2019attack}.

\subsection{Address translation and TLB}
Virtual memory systems abstract physical memory to achieve process isolation and efficient memory management.
In this system, each process operates with its own independent virtual address.
When a memory access occurs, these virtual addresses must be translated into their corresponding physical memory address.
This crucial task is typically managed by the Memory Management Unit (MMU), a key hardware component that mediates the interaction between the CPU and the physical memory system.
The MMU accomplishes this translation using a process-specific mapping table, commonly known as the page table.
However, because the page table is stored in main memory, the process of translating addresses via page table walk can be time-consuming.

To address this, modern MMUs include the Translation Lookaside Buffer (TLB), a specialized form of cache memory designed to accelerate the address translation process.
The TLB retains the most recently used virtual-to-physical address mappings, enabling swift address translation without repetitive page table walks for the same address.
When accessing memory, if the target address is already cached in the TLB, (i.e., TLB hit), the time required for address translation is substantially reduced. 
Conversely, if the target address is not present in the TLB, (i.e., TLB miss), address translation necessitates resorting to the conventional page table walk. 

\subsection{Cache-based side-channel attack}
Caches are designed to reduce memory access latency by leveraging temporal and spatial locality.
However, they have become a source of side-channel attacks~\cite{yarom2014flush+,gruss2016flush+,liu2015last,irazoqui2015s,zhao2024last, yan2019attack, purnal2021prime+,chen2024prefetchx, shusterman2021prime+} because cache state influences timing information, which can be exploited to extract security-sensitive data.
To achieve this, an attacker should be able to manipulate the cache state into a known state and monitor the state transition by measuring differences in the cache access latency.

For this purpose, three types of attack are typically employed.
The \textit{flush}-based attack~\cite{yarom2014flush+, gruss2016flush+,gulmezouglu2015faster} uses the \texttt{clflush} instruction to flush a target cache line, and observes the change in the cache state by using timing information.
While this method is straightforward, it relies on both \texttt{clflush} and memory sharing between the attacker and the victim.
To overcome the dependency on the \texttt{clflush}, the \textit{evict}-based attack~\cite{gruss2015cache} evicts the desired cache line using an eviction set \textit{i.e.,} a collection of addresses mapped to the same cache set as the target address.
The third, \textit{prime}-based attack~\cite{liu2015last, irazoqui2015s, shusterman2021prime+}, does not require both the \texttt{clflush} and memory sharing.
For this, the attacker primes the target set with the eviction set and probes the state transitions of the primed cache lines.

The LLC has been commonly used for cross-core attacks because it is shared among processor cores. 
However, it is ineffective for non-inclusive LLC, as the victim's memory access only affects its private cache.
As a result, attackers must instead target the directory structure, which maintains inclusivity for the private cache~\cite{yan2019attack, purnal2021prime+, zhao2024last}.

\subsection{New ISA extensions in Sapphire Rapids}
Since the launch of the 4th Gen Intel Xeon Scalable Processors, codenamed Sapphire Rapids, Intel has introduced several advancements to better support workloads in data centers~\cite{Intel2021archi}.
These workloads primarily focus on specialized tasks, such as AI, networking, and in-memory database analysis.
Given that such tasks are not suitable for executing efficiently on general-purpose processors, Intel has developed new accelerators, including the Intel Dynamic Load Balancer (Intel DLB)~\cite{Intel2020DLB1, Intel2020DLB2}, Intel Data Streaming Accelerator (Intel DSA)~\cite{Intel2020DSA}, Intel In-Memory Analytics Accelerator (Intel IAA)~\cite{Intel2020IAA}, and Intel Advanced Matrix Extensions (Intel AMX)~\cite{Intel2020AMX}, to offload these tasks onto dedicated hardware.
To support their usage, Intel adds new ISA extensions and Intel Accelerator Interfacing Architecture (Intel AIA) to provide an interface between software and these accelerators.


In addition, Intel has introduced a new ISA extension to address bottlenecks in multi-threaded, large-scale workloads.
These bottlenecks often occur when requested data is located in another core's private cache rather than in the shared LLC, leading to low access latency.
To optimize cache performance and reduce this delay, Intel has implemented a new instruction called the \texttt{cldemote}.
It is designed to move data from a private cache to a shared LLC~\cite{Intel2023SDM, Intel2024AORM, Intel2024IASEFF}.
By demoting data to the LLC, the usage of \texttt{cldemote} can enhance the overall efficiency of multi-threaded workloads.

\section{Analysis of CLDEMOTE Instruction}
\label{sec:key_property}
In this section, we analyze architectural and microarchitectural properties of the \texttt{cldemote} instruction.
We first examine its architectural characteristics (Section~\ref{subsec:cldemote}), and then analyze its execution latency with respect to the location of the target address in the cache hierarchy (Section~\ref{subsec:Timing}).
Finally, we investigate its property regarding fault suppression (Section~\ref{subsec:fault_suppression}).

\noindent\textbf{Experimental setup.}
For the analysis, we conduct experiments on an Intel Xeon Silver 4510T processor, codenamed Sapphire Rapids, operating with Ubuntu 24.04 (kernel version 6.8.0-38).
The processor features a non-inclusive LLC.
Since the release of Linux kernel 5.5, 5-level paging for the x86\_64 architecture has been enabled by default~\cite{Linux20195level,Kernelnewbies20205level}.
Our experiments utilize the default configuration, which supports a 56-bit virtual address space.
We also enable the Simultaneous Multi-Threading (SMT) on our processor.
Throughout this paper, we use this experimental setup unless specified otherwise.

\begin{figure}[t!]
    \centerline{\includegraphics[scale=0.28]{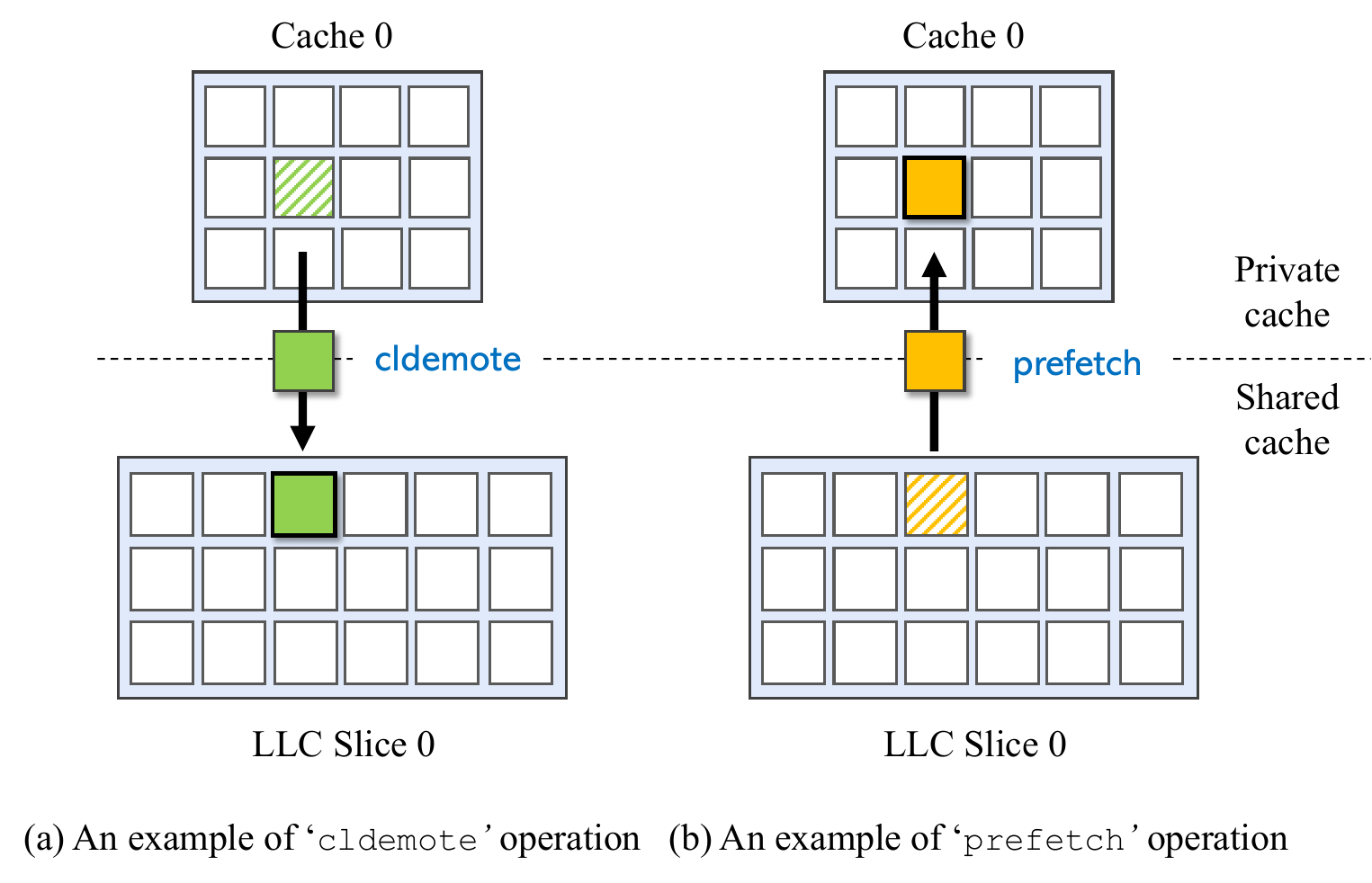}}
    \caption{Comparison of \texttt{cldemote} and \texttt{prefetch} behavior.}
    \label{fig:fig1}
    \vspace{-0.4cm}
\end{figure}

\subsection{Cache line demote operation}
\label{subsec:cldemote}
The primary function of \texttt{cldemote} is to provide a hardware hint that moves (or demotes) data from a private cache to a shared LLC. 
This operation contrasts with prefetch instructions (e.g., the \texttt{prefetchX} family in x86), which bring data closer to the core, as illustrated in Figure~\ref{fig:fig1}.

\texttt{cldemote} was introduced through Intel Architecture Communications Technology~\cite{Intel2024Lab} to address performance bottlenecks in multi-core systems. 
In producer-consumer applications~\cite{Intel2024AORM}, frequent load/store operations often trigger expensive cross-core snooping to retrieve data from other cores’ private caches. 
By demoting cache lines to the LLC, \texttt{cldemote} avoids such costly snooping overhead.
Specifically, a core holding the data in its private cache can facilitate inter-core data transfers by placing the data in the LLC, thereby ensuring more efficient retrieval when other cores request the same data.

Like other cache management instructions (e.g., \texttt{clflush}, \texttt{clflushopt}, and \texttt{clwb}), the \texttt{cldemote} instruction is available at all privilege levels. 
Its unique architectural property is that it does not write back modified cache lines to main memory; instead, it moves them to the LLC while preserving the same coherence state. 
For additional details on checking processor support for \texttt{cldemote}, refer to Appendix~A.

Below are two identified properties of the operation.

\begin{tcolorbox}[boxsep=1pt,left=5pt,right=5pt,top=5pt,bottom=5pt]
\textbf{P1. Inter-cache state transition.}
The execution of \texttt{cldemote} moves a cache line from a private cache to a shared LLC.

\textbf{P2. Unprivileged access.}
\texttt{cldemote} is accessible to an unprivileged user.
\end{tcolorbox}

\subsection{Execution latency}
\label{subsec:Timing}

Given the nature of \texttt{cldemote}, we hypothesize that its execution time depends on where the data resides within the cache hierarchy. 
To verify this, we designed an experiment measuring the instruction’s execution delay under different cache locations.

In our setup, we consider four possible locations for the target address $\mathcal{T}$: L1d, L2, the LLC, and main memory. 
For each case, we measure the execution latency 1{,}000{,}000 times using the \texttt{rdtsc} instruction to obtain accurate average timing information.

\begin{figure}[t!]
    \centerline{\includegraphics[scale=0.16]{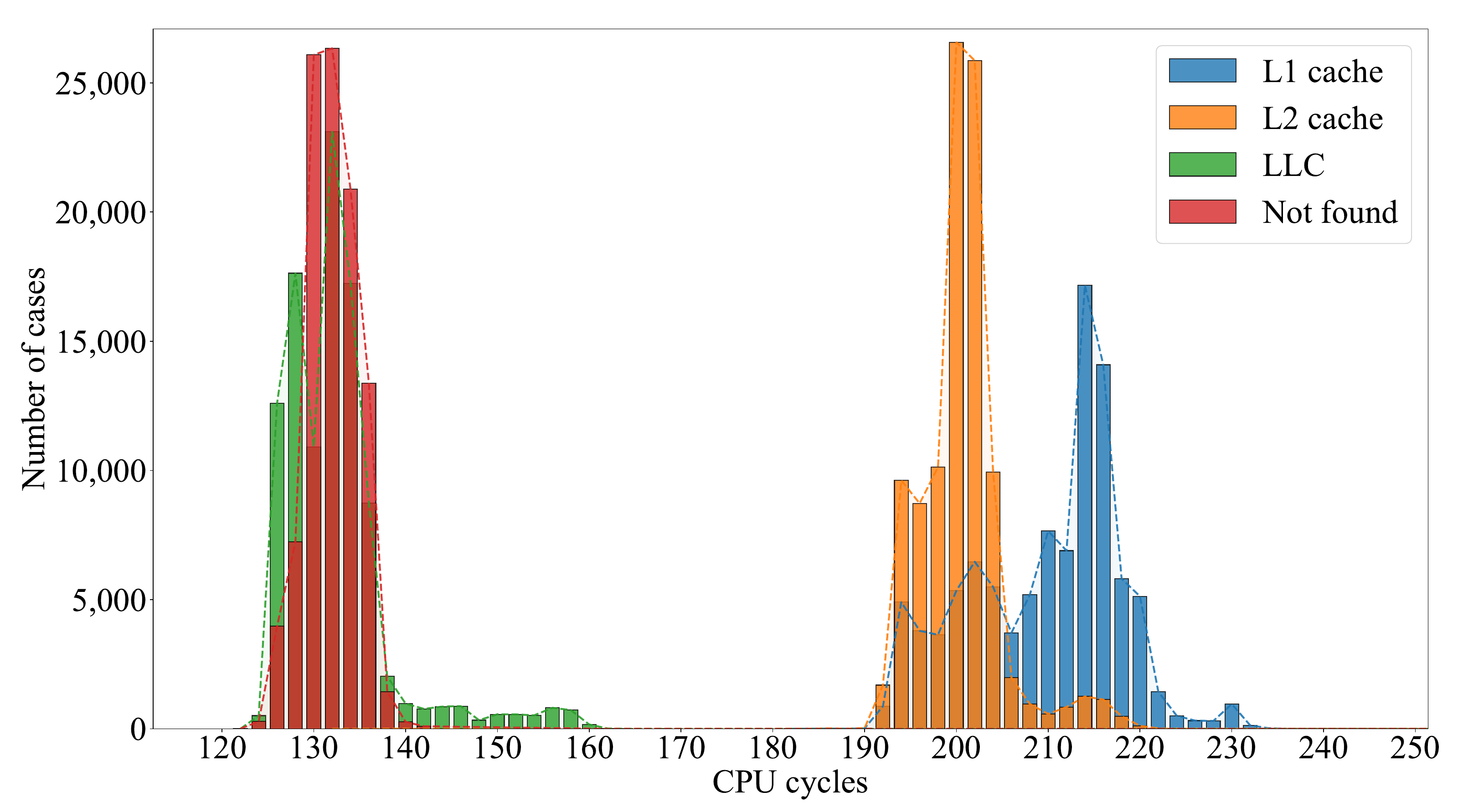}}
    \caption{Execution latency of \texttt{cldemote} instruction on various cache levels.}
    \label{fig:fig2}
    \vspace{-0.4cm}
\end{figure}

Figure~\ref{fig:fig2} shows the distribution of execution latencies for these scenarios. 
We observe that the latency increases when $\mathcal{T}$ is located closer to the core within the private cache. 
Specifically, the average latencies for L1d and L2 are 210 cycles and 200 cycles, respectively. 
If $\mathcal{T}$ is not present in the private cache (i.e., it resides in the shared LLC or is not found in any cache), the latency is around 132 cycles. 
In addition, we tested \texttt{cldemote} on L1i to assess its ability to demote cache lines containing instructions; our results confirm that it works for the instruction cache as well.

We also observe that \texttt{cldemote} moves the cache line to the LLC only if it currently resides in the private cache. 
This finding aligns with Intel’s documentation, which states that \textit{“If the line is not found in the cache, the instruction will be treated as a NOP.”}~\cite{Intel2023SDM}

\begin{tcolorbox}[boxsep=1pt,left=5pt,right=5pt,top=5pt,bottom=5pt]
\textbf{P3. Timing difference.}
The execution time of \texttt{cldemote} can be used to infer the state of a cache line located in private caches (i.e., L1 and L2).
\end{tcolorbox}

\subsection{Fault suppression}
\label{subsec:fault_suppression}
Intel’s documentation~\cite{Intel2024IASEFF} states that \texttt{cldemote} does not trigger a page fault even if a memory access violates permissions. 
To verify this claim, we conducted an additional experiment under our existing setup.
Specifically, we examined three scenarios in which a user-initiated memory reference would normally result in a page fault: 1) dereferencing an invalid address in user space, 2) dereferencing a valid address in kernel space, and 3) dereferencing an invalid address in kernel space.
For each scenario, we issued the \texttt{cldemote} instruction on the faulting address and checked whether it caused a fault.

Our results show that no faults were triggered in any of these scenarios, indicating that \texttt{cldemote} does not raise exceptions even if the demoted memory address violates access permissions. 
This fault-suppressing property allows for more effective and practical side-channel attacks, as seen in previous work leveraging similar behavior in other extensions such as TSX~\cite{jang2016breaking}, software-based prefetching~\cite{gruss2016prefetch, lipp2022prefetch}, and AVX~\cite{choi2023avx, kim2023avx}.

\begin{tcolorbox}[boxsep=1pt,left=5pt,right=5pt,top=5pt,bottom=5pt]
\textbf{P4. Fault suppression}.
\texttt{cldemote} has fault suppression capability. Therefore, it does not trigger fault even if the memory address being accessed violates access permissions.
\end{tcolorbox}

\section{Case Study 1 - Side-Channel Attacks}
\label{sec:case_study1}

We present two side-channel attacks based on \texttt{cldemote}-based attack primitives.
First, we build two novel attack primitives, Flush+Demote and Demote+Time, that exploit the properties of \texttt{cldemote} that we revealed in the previous section (Section~\ref{sec:attack_primitives}).
Then, we present an attack that constructs a covert channel between sibling cores by using Flush+Demote (Section~\ref{subsec:covert_channel}).
We also present another attack that derandomizes the kernel base address with Demote+Time (Section~\ref{subsec:breaking_kaslr}).

\subsection{Building attack primitives}
\label{sec:attack_primitives}


\subsubsection{Flush+Demote}
\label{subsec:flush+demote}
Flush+Demote aims to infer a victim's memory access pattern on the target address $\mathcal{T}$ by observing a state change in the CPU cache.
To implement this, we leverage three architectural properties of \texttt{cldemote}, \textbf{P1}, \textbf{P2} and \textbf{P3}, which are identified in the previous section.
Flush+Demote has the same requirement as Flush+Reload~\cite{yarom2014flush+}, where the target address $\mathcal{T}$ is shared with a victim.
This attack has the following three steps.

\vspace{-0.1cm}
\begin{enumerate}[Step 1.]
\setlength{\itemsep}{-0.20em}
\item (\textit{Flush}) It flushes $\mathcal{T}$ from the entire cache.
\item (\textit{Wait}) It waits for the victim's load operation, which may depend on the victim's secret.
\item (\textit{Demote}) It executes \texttt{cldemote} on $\mathcal{T}$ and then measures its execution latency.
If the time exceeds a predefined threshold, it means that the victim has made a memory access to $\mathcal{T}$; otherwise, it does not.
\end{enumerate}
\vspace{-0.1cm}

\begin{algorithm}[t!]
\DontPrintSemicolon
  \SetKwFunction{Fattack}{\textit{attacker\_thread}}
  \SetKwFunction{Fvictim}{\textit{victim\_thread}}
  
  \KwInput{A target address \textit{addr} }
  \KwOutput{none}

  \SetKwProg{Fn}{procedure}{:}{end}
  \Fn{\Fattack{\textit{addr}}}{
        \For{\textit{i} $\leftarrow1$ \KwTo $100\;000$}
            {
              \textit{clflush(\textit{addr})}\;
              \textit{wait\_thread()}\;
               start $\leftarrow$ \textit{rdtsc()}\;
               \textit{cldemote(\textit{addr})}\;
               \textit{mfence()}\;
               end $\leftarrow$ \textit{rdtsc()}\;
            }
  }
  
 \SetKwProg{Fn}{procedure}{:}{end}
  \Fn{\Fvictim{\textit{addr}}}{
                
        \For{\textit{i} $\leftarrow1$ \KwTo $100\;000$}
            {
               \If{\textit{i} \% 2 is 0}{
                  \textit{maccess(\textit{addr})}
               }
               \textit{wait\_thread()}\;
            }
  }
\caption{Flush+Demote attack primitive}
\label{algorithm:algo1}
\end{algorithm}

\noindent\textbf{Evaluation.}
%
To demonstrate the effectiveness of Flush+Demote, we conduct an experiment based on the methodology proposed by Guo et al.~\cite{guo2022adversarial}. 
We begin by preparing a target address $\mathcal{T}$ with read-only permissions using \texttt{mmap()} and the \texttt{MAP\_PRIVATE} and \texttt{MAP\_ANONYMOUS} flags. 
Next, we create two threads running Algorithm~\ref{algorithm:algo1} on $\mathcal{T}$—an attacker thread and a victim thread—pinned to sibling logical cores within the same physical core.

As shown in the algorithm, the attacker thread first flushes $\mathcal{T}$ and then waits for the victim to execute (lines~3--4). 
We synchronize these two concurrent threads using a mutex. 
The victim thread performs memory accesses to $\mathcal{T}$ only on even iterations, while doing nothing on odd iterations (lines~13--15). 
Finally, the attacker measures the execution latency of \texttt{cldemote} 100{,}000 times (lines~5--8).

\begin{table}[]
\centering
\renewcommand{\arraystretch}{1.1}
\caption{Comparison of memory sharing-based attacks.}
\label{table:ttttt}
\begin{adjustbox}{width=1\columnwidth}
\begin{tabular}{cccccc}
\Xhline{3\arrayrulewidth}
\textbf{Primitive}                         & \begin{tabular}[c]{@{}c@{}}\textbf{Hit}\\ \textbf{(cycles)}\end{tabular} & \begin{tabular}[c]{@{}c@{}}\textbf{Miss}\\ \textbf{(cycles)}\end{tabular} & \begin{tabular}[c]{@{}c@{}}\textbf{Timing diff.} \\ \textbf{(cycles)}\end{tabular} & \begin{tabular}[c]{@{}c@{}}\textbf{Recovery rate of}\\ \textbf{RSA private key}\end{tabular} & \begin{tabular}[c]{@{}c@{}}\textbf{Leakage rate}\\ \textbf{in Spectre v1}\end{tabular} \\ \hline\hline
Stream+Reload~\cite{weber2021osiris}     &   61         &    309        &     248    &  N/A    &  17.98 KB/s          \\
Flush+Reload~\cite{yarom2014flush+}       &   58         &    308        &     250    &  99.4\%  &  15.98 KB/s          \\
Flush+Flush~\cite{gruss2016flush+}        &   213        &    139        &     74     &  99.4\%  &  24.81 KB/s          \\
Flush+Demote                              &   208        &    121        &     94     &  98.6\%   &  26.68 KB/s          \\ \Xhline{3\arrayrulewidth}
\end{tabular}
\end{adjustbox}
\vspace{-0.4cm}
\end{table}

Table~\ref{table:ttttt} presents our experimental results, providing average timing information for each side-channel primitive based on different cache states and their respective timing differences. 
For comparison, we also evaluated the performance of other memory-sharing-based side-channel attacks.

Flush+Reload~\cite{yarom2014flush+} and Stream+Reload~\cite{weber2021osiris} exhibit the largest timing difference between cache hits and misses (approximately 250 cycles). 
In contrast, Flush+Demote has a reduced difference of 94 cycles, and Flush+Flush shows the smallest gap. These findings suggest that timing information is significantly influenced by the instruction used to measure execution latency.

Furthermore, we observe that the execution latency of \texttt{cldemote} is slightly lower than \texttt{clflush}, even when no cache line is present to demote or flush. 
This is because \texttt{clflush} must check the LLC to remove a cache line from the hierarchy, whereas \texttt{cldemote} does not.

We also conducted two additional experiments to evaluate how a victim’s memory access patterns influence the capability of the primitive:
1) recovering RSA secret key from GnuPG 1.4.13, and 2) measuring the leakage rate in Spectre v1.
In the first experiment, we attacked the square-and-multiply exponentiation in RSA (GnuPG 1.4.13) by probing \texttt{mpih\_sqr\_n()} (square), \texttt{mpihelp\_divrem()} (reduce), and \texttt{mpih\_n()} (multiply). 
We used a 1{,}200-cycle time slot to recover 512 bits of a 1{,}024-bit secret key, achieving a 99\% success rate across all primitives. 
However, Stream+Reload was ineffective on read-only pages (e.g., shared libraries).

In the second experiment, we implemented covert channel attacks in Spectre v1, replacing the traditional Flush+Reload with alternative cache attack primitives. 
Our results show that Stream+Reload achieves a leakage rate of 17.98~KB/s, outperforming Flush+Reload. 
This is because \texttt{clflush} sets the cache line to the invalid state, while \texttt{movnt} marks it as modified. 
Consequently, a subsequent load instruction must transition from invalid to modified, incurring higher latency than transitioning from modified to another state.

By contrast, Flush+Flush and Flush+Demote attain leakage rates of 24.81~KB/s and 26.68~KB/s, respectively—a 48\% improvement over both Stream+Reload and Flush+Reload. 
Their ability to preserve the cache state before and after measuring execution latency makes them more effective, particularly when the victim’s behavior causes numerous cache misses.

\noindent\textbf{Cross-core attack with Flush+Demote.}
While we have demonstrated the effectiveness of the Flush+Demote across SMT threads, the primitive cannot be used for a cross-core cache attack scenario.
This is because \texttt{cldemote} can only demote cache lines within the sibling core where it executes.
Therefore, if $\mathcal{T}$ resides on a different physical core, an attacker from another core cannot demote the cache line to the shared LLC.
As a result, the attacker is unable to infer the victim's memory access patterns in a cross-core attack scenario.

\subsubsection{Demote+Time}
\label{subsec:demote+time}

Our second attack primitive distinguishes valid (physically backed) kernel addresses from invalid ones by measuring the execution time of \texttt{cldemote}. 
We refer to this technique as Demote+Time, which leverages specific TLB behavior on Intel CPUs. 
In particular, the CPU can allocate a TLB entry for an address that is valid yet inaccessible from user space, resulting in a faster execution time due to TLB hits. 
Conversely, an invalid address incurs a page table walk, leading to longer execution times. 
Since \texttt{cldemote} does not trigger faults (the property \textbf{P4}), it can be applied to otherwise inaccessible kernel addresses without requiring additional fault handling or suppression. 

This attack primitive proceeds in two steps:

\vspace{-0.1cm}
\begin{enumerate}[Step 1.]
\setlength{\itemsep}{-0.20em}
\item (\textit{Demote}) It executes \texttt{cldemote} on the target kernel address $\mathcal{T}$.

\item (\textit{Time}) It measures its execution latency. 
\end{enumerate}
\vspace{-0.1cm}

\noindent\textbf{Evaluation.}
To demonstrate the effectiveness of Demote+Time in differentiating valid addresses from invalid ones, we designed an experiment and compared our results with software-based prefetch instructions~\cite{gruss2016prefetch, lipp2022prefetch}. 
We prepare a target page $\mathcal{P}$ and issue the \texttt{cldemote} instruction on it, measuring execution latency one million times.

We consider two cases, each corresponding to a different TLB state: TLB hit and TLB miss.
To induce a TLB hit, we execute \texttt{cldemote} twice in succession. 
The first execution loads the address translation into the TLB, ensuring that the subsequent execution experiences a TLB hit. 
For the TLB miss case, we flush the TLB prior to each execution. 
We repeat this procedure with \texttt{prefetch} for comparison with \texttt{cldemote}.

\begin{figure}[t!]
    \centerline{\includegraphics[scale=0.31]{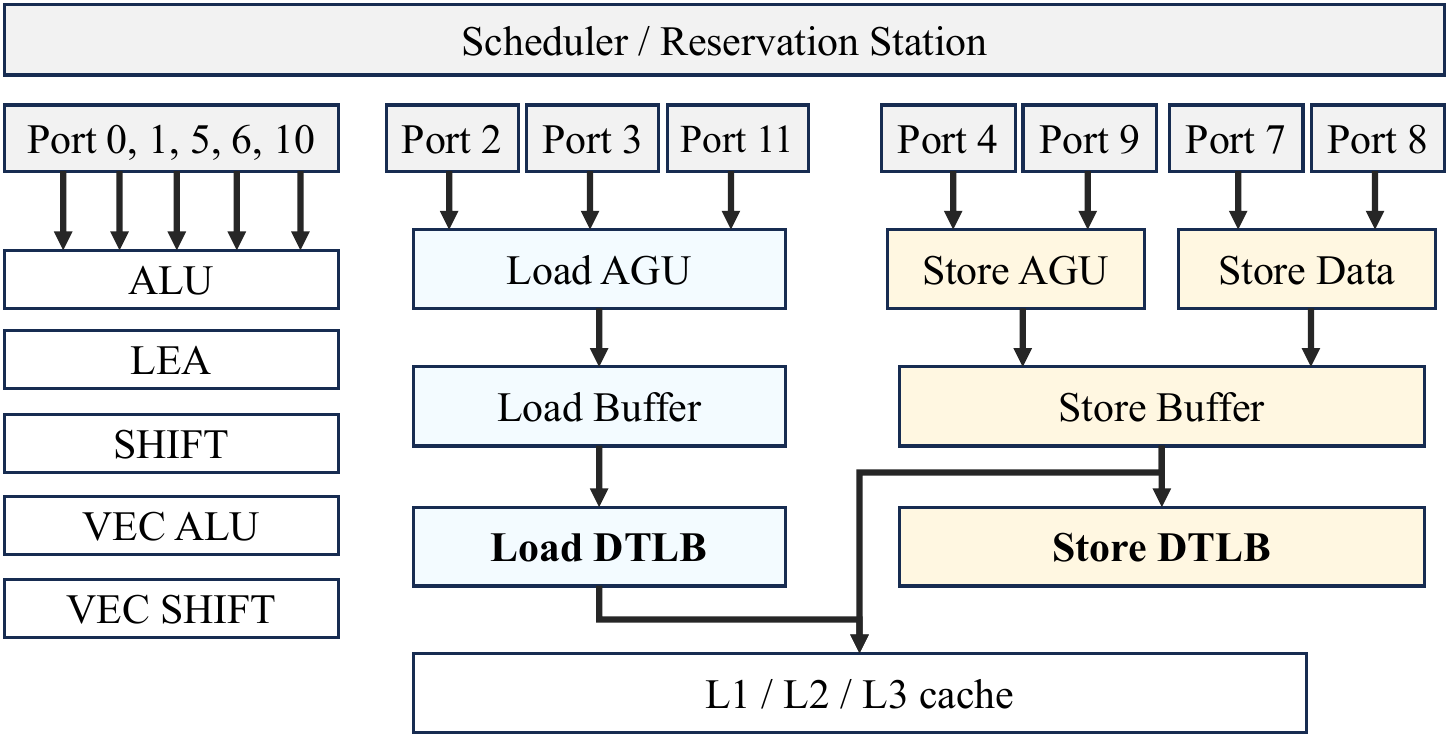}}
    \caption{Overview of Golden Cove microarchitecture.}
    \label{fig:fig5}
    \vspace{-0.5cm}
\end{figure}

\begin{table*}[t!]
\tiny
\centering
\renewcommand{\arraystretch}{1.0}
\caption{The execution time of \texttt{prefetch} and \texttt{cldemote} on memory pages with various permission bits and memory types.  }
\label{table:table4}
\begin{adjustbox}{width=0.99\textwidth}
\begin{tabular}{ccccc|c|cc|cc}
\Xhline{2\arrayrulewidth}
\multicolumn{5}{c|}{\textbf{Permission bits$^{*}$}} & \textbf{Memory types}     & \multicolumn{2}{c|}{\texttt{prefetch}} & \multicolumn{2}{c}{\texttt{cldemote}}  \\ \hline
\textbf{P}  & \textbf{U} & \textbf{D} & \textbf{NX} & \textbf{A} &                 & \textbf{TLB hit}      & \textbf{TLB miss}     & \textbf{TLB hit}      & \textbf{TLB miss}     \\ \hline\hline
\CIRCLE     & \CIRCLE    & \CIRCLE    & \CIRCLE     & \CIRCLE    & Write-back      & 113 ($\sigma=\:$6.08) & 133 ($\sigma=\:$7.76) & 160 ($\sigma=\:$6.42) & 180 ($\sigma=\:$6.07) \\
\CIRCLE     & \Circle    & \CIRCLE    & \CIRCLE     & \Circle    & Write-back      & 132 ($\sigma=\:$5.92) & 132 ($\sigma=\:$7.46) & 137 ($\sigma=\:$6.78) & 136 ($\sigma=\:$5.30) \\
\CIRCLE     & \CIRCLE    & \CIRCLE    & \CIRCLE     & \Circle    & Write-back      & 132 ($\sigma=\:$6.94) & 133 ($\sigma=\:$7.00) & 137 ($\sigma=\:$6.54) & 136 ($\sigma=\:$4.78) \\
\CIRCLE     & \Circle    & \CIRCLE    & \CIRCLE     & \CIRCLE    & Write-back      & 112 ($\sigma=\:$5.04) & 132 ($\sigma=\:$7.37) & 114 ($\sigma=\:$5.38) & 136 ($\sigma=\:$4.87) \\
\Circle     & \CIRCLE    & \CIRCLE    & \CIRCLE     & \CIRCLE    & Write-back      & 115 ($\sigma=\:$5.23) & 133 ($\sigma=\:$7.00) & 139 ($\sigma=\:$7.09) & 148 ($\sigma=\:$6.39) \\
\Circle     & \Circle    & \CIRCLE    & \CIRCLE     & \CIRCLE    & Write-back      & 115 ($\sigma=\:$5.76) & 132 ($\sigma=\:$6.45) & 138 ($\sigma=\:$6.24) & 149 ($\sigma=\:$7.88)  \\
\CIRCLE     & \CIRCLE    & \CIRCLE    & \Circle     & \CIRCLE    & Write-back      & 113 ($\sigma=\:$5.92) & 132 ($\sigma=\:$6.32) & 160 ($\sigma=\:$5.86) & 180 ($\sigma=\:$6.10)  \\
\CIRCLE     & \CIRCLE    & \Circle    & \CIRCLE     & \CIRCLE    & Write-back      & 112 ($\sigma=\:$5.68) & 134 ($\sigma=\:$7.98) & 160 ($\sigma=\:$6.38) & 180 ($\sigma=\:$5.14)  \\
\Circle     & \Circle    & \CIRCLE    & \CIRCLE     & \Circle    & Write-back      & 132 ($\sigma=\:$5.96) & 132 ($\sigma=\:$7.05) & 148 ($\sigma=\:$6.99) & 148 ($\sigma=\:$6.48) \\ 
\CIRCLE     & \CIRCLE    & \CIRCLE    & \CIRCLE     & \CIRCLE    & Write-protected & 113 ($\sigma=\:$6.04) & 132 ($\sigma=\:$7.13) & 114 ($\sigma=\:$5.07) & 137 ($\sigma=\:$7.03) \\  
\CIRCLE     & \CIRCLE    & \CIRCLE    & \CIRCLE     & \CIRCLE    & Uncacheable     & 113 ($\sigma=\:$5.79) & 135 ($\sigma=\:$8.31) & 115 ($\sigma=\:$6.06) & 138 ($\sigma=\:$7.92) \\ \Xhline{2\arrayrulewidth}
\end{tabular}
\end{adjustbox}
\footnotesize{\begin{flushleft} $^{*}$ \textbf{P} denotes Present bit, \textbf{U} denotes User/Supervisor bit, \textbf{D} denotes Dirty bit, \textbf{NX} denotes Non eXecutable bit, and \textbf{A} denotes Accessed bit. \\ The symbol (\CIRCLE) indicates that the bit is set to 1, and the symbol (\Circle) indicates that the bit is set to 0.    \end{flushleft} }
\vspace{-0.45cm}
\end{table*}

To perform a comprehensive evaluation, we tested various pages with different permission bits and memory types.  
Table~\ref{table:table4} presents our experimental results using \texttt{cldemote} and \texttt{prefetch}.  
As expected, both instructions exhibit longer execution times during TLB misses than TLB hits.  

However, when the \texttt{A} bit in the accessed page table entry is set to 0, we observe that TLB hits and misses incur similar execution times.  
This behavior arises from two architectural features of Intel processors:  
first, the processor caches a virtual-to-physical address pair only if the \texttt{A} bit is set to 1 in every page table entry involved in address translation~\cite{Intel2023SDMv3};  
second, neither \texttt{cldemote} nor \texttt{prefetch} sets the \texttt{A} bit to 1 in the page table entry.  

We also found that \texttt{cldemote}’s execution time is influenced by both the \texttt{P} bit and the \texttt{U} bit, whereas \texttt{prefetch} is not.  
Specifically, when only the \texttt{P} bit is set to 1, \texttt{cldemote} takes 114 cycles on a TLB hit and 136 cycles on a TLB miss.  
When only the \texttt{U} bit is set to 1, the execution time increases to 139 cycles for a TLB hit and 148 cycles for a TLB miss.  
Furthermore, when both the \texttt{P} and \texttt{U} bits are set to 1, \texttt{cldemote} exhibits its longest execution times, reaching 160 cycles on a TLB hit and 180 cycles on a TLB miss.  

This result indicates that the time required to suppress a fault rises when the page is marked as non-present or is located in kernel address space.  
Notably, unlike \texttt{prefetch}, the execution time of \texttt{cldemote} can reveal both the TLB state and the permission bits in the page table entry.  

Additionally, Intel’s documentation states that \emph{“performing \texttt{cldemote} on an uncacheable type of page may cause the instruction to be ignored”}~\cite{Intel2023SDM}.  
Nevertheless, our observations show that, even if the execution is ignored on uncacheable memory, the processor still records the resolved virtual-to-physical address pair in the TLB.

\begin{table}[t!]
\centering
\renewcommand{\arraystretch}{1.1}
\caption{The number of measured TLB miss events when executing the \texttt{cldemote} and \texttt{prefetch} 1 million times.}
\label{table:table3}
\begin{adjustbox}{width=1\columnwidth}
\begin{tabular}{cccc}
\Xhline{3\arrayrulewidth}
\textbf{Instruction}              & \textbf{Address type}             & \textbf{$dTLB_{load}$ misses} & \textbf{$dTLB_{store}$ misses} \\ \hline\hline
\multirow{2}{*}{\texttt{cldemote}} & Valid addr   &  0                                  &  0                                   \\
                                   & Invalid addr &  0                                  &  2,000,000                                   \\ \hline
\multirow{2}{*}{\texttt{prefetch}} & Valid addr   &  0                                  &  0                                   \\
                                   & Invalid addr &  0                                  &  0                                   \\ \Xhline{3\arrayrulewidth}
\end{tabular}
\end{adjustbox}
\vspace{-0.5cm}
\end{table}

\noindent\textbf{An in-depth analysis of the TLB in Golden Cove.}
To analyze the interaction between TLB and memory operations such as \texttt{cldemote} and \texttt{prefetch}, we examine the microarchitectural properties of the Sapphire Rapids processor, which is based on the Golden Cove architecture.  
According to Intel's documentation~\cite{Intel2024AORM}, Golden Cove introduces several microarchitectural modifications, including changes to the TLB structure.  
Specifically, Golden Cove separates data TLBs into two types: one for load operations ($dTLB_{load}$) and another for store operations ($dTLB_{store}$), as illustrated in Figure~\ref{fig:fig5}.  

To investigate the architectural properties of these TLBs, we conduct an analysis using performance monitoring events.  
We prepare two addresses: one is a valid kernel address and the other is an invalid kernel address.  
We then execute \texttt{cldemote} and \texttt{prefetch} one million times for each address, measuring the change in the $dTLB_{load}$ miss events (\texttt{DTLB\_LOAD\_MISSES.WALK\_COMPLETED}) and $dTLB_{store}$ miss events (\texttt{DTLB\_STORE\_MISSES.WALK\_COMPLETED}).  
Since \texttt{cldemote} is a store-type memory operation and \texttt{prefetch} is a load-type memory operation, they utilize different data TLBs respectively.  

The experimental results are shown in Table~\ref{table:table3}.  
Our findings indicate that in Golden Cove, $dTLB_{load}$ caches virtual-to-physical address translations even when the kernel address is invalid (i.e., not present in memory).  
In contrast, $dTLB_{store}$ updates the TLB only when the address is valid.  
We also observed that executing \texttt{cldemote} with an invalid address triggers two page table walks.  
This behavior aligns with observations on AMD's Zen microarchitecture, where issuing a \texttt{prefetch} to an invalid address also results in two page table walks~\cite{lipp2022prefetch}.  

Executing \texttt{prefetch} on an inaccessible kernel address allocates a TLB entry, regardless of its validity.  
However, the execution time differs slightly between valid and invalid addresses, even though both trigger TLB hits.  
We measure 108 cycles for a valid address and 112 cycles for an invalid address.  
This result suggests that handling a TLB hit on an invalid address requires extra time because there is no physically backed memory.  
In addition, Demote+Time enables an unprivileged attacker to distinguish the page table level by measuring the time required for the page table walk.  
We provide further details in Appendix~B.

\subsection{Building covert-channel}
\label{subsec:covert_channel}

\noindent\textbf{Threat model.}
We consider two unprivileged processes, referred to as the sender and the receiver.  
These processes reside on the same physical core but execute on different logical cores (i.e., sibling threads).  
They aim to communicate to exfiltrate security-sensitive information, circumventing the software-based security boundary.  
However, because they remain strictly isolated, direct communication between them is not feasible.

\noindent\textbf{Constructing covert channel.}
To establish a communication channel between the sender and receiver, they decided to leverage the CPU cache for the channel.
To achieve this, they construct the channel based on the Flush+Demote.
This communication protocol is performed through the following four steps.

\vspace{-0.1cm}
\begin{enumerate}[Step 1.]
\setlength{\itemsep}{-0.20em}
\item (\textit{Agreement}) 
The sender and receiver agree on a cache line $\mathcal{T}$, which will serve as a covert channel for message transmission.
This can be achieved by sharing a memory page between them by mapping the same shared library.
They also agree on the predefined time window $\mathcal{W}$, which is the time it takes to transfer 1-bit data.

\item (\textit{Synchronization})
The sender and receiver synchronize the channel based on the time stamp counter.

\item (\textit{Send})
To transmit a message, the sender encodes data within the cache.
To send a bit `1’, the sender repeatedly accesses memory to allocate the cache line $\mathcal{T}$ during the $\mathcal{W}$.
To send a bit `0’, the sender remains idle during the corresponding time window.

\item (\textit{Receive}) 
The receiver decodes the transmitted messages by performing Flush+Demote once within the $\mathcal{W}$.
If the execution time of the \texttt{cldemote} exceeds a predefined threshold, it indicates that the sender transmitted a bit value of `1'.
Otherwise, it indicates a bit value of `0'.
\end{enumerate}
\vspace{-0.1cm}

\begin{table}[t!]
\centering
\renewcommand{\arraystretch}{1.1}
\caption{The maximum channel capacity and its corresponding bit error possibility among flush-based primitives.}
\label{table:table5}
\begin{adjustbox}{width=0.9\columnwidth}
\begin{tabular}{ccc}
\Xhline{3\arrayrulewidth}
             & \textbf{Channel capacity} & \textbf{Bit error possibility} \\ \hline\hline
Flush+Reload~\cite{yarom2014flush+} & 1.248 Mbps       & 0.015\%                      \\
Flush+Flush ~\cite{gruss2016flush+} & 2.848 Mbps       & 0.022\%                      \\
Flush+Demote                        & 2.849 Mbps       & 0.018\%                      \\ \Xhline{3\arrayrulewidth}
\end{tabular}
\end{adjustbox}
\vspace{-0.2cm}
\end{table}

\noindent\textbf{Evaluation.}
For simplicity, we construct a naive covert channel without implementing error-correction codes or other performance-optimization techniques.  
For the evaluation, we transmitted a 1,000,000-bit message with an equal distribution of bits `1’ and `0’.  
By varying the size of the time window $\mathcal{W}$, we assess the channel’s performance in terms of channel capacity and bit error rate, which are common metrics for evaluating covert channels~\cite{okhravi2010design, pessl2016drama, paccagnella2021lord, guo2022adversarial}.  

\begin{figure}[t!]
    \centerline{\includegraphics[scale=0.18]{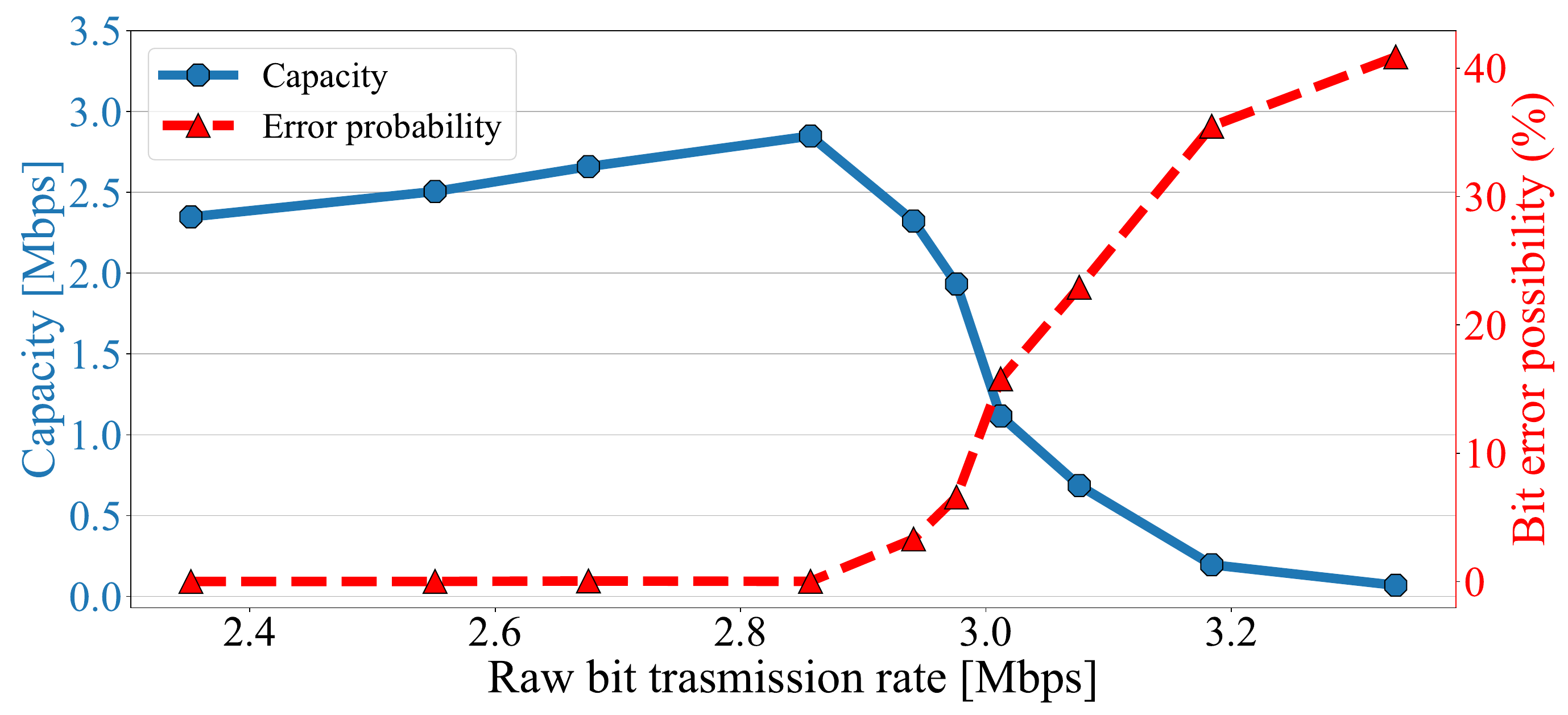}}
    \caption{The channel capacity and bit error possibility for Flush+Demote covert channel.}
    \label{fig:fig6}
    \vspace{-0.4cm}
\end{figure}

Figure~\ref{fig:fig6} and Table~\ref{table:table5} illustrate the performance of our Flush+Demote covert channel and compare it with prior work by reproducing their channels~\cite{yarom2014flush+,gruss2016flush+}.  
Flush+Demote shows an increase in capacity up to a raw bit transmission rate of 2.857~Mbps.  
At this rate, the channel achieves its maximum capacity with a bit error rate of 0.018\%.  
Under the same experimental setup, Flush+Reload attains roughly half the capacity of Flush+Demote, yet has a lower bit error rate (see Figure~\ref{fig:fig_a_cc2}).  
In contrast, Flush+Flush exhibits a capacity similar to Flush+Demote but suffers from a higher bit error rate (see Figure~\ref{fig:fig_a_cc1}).  
We attribute these performance differences to the fundamental properties of each primitive, as detailed in Table~\ref{table:ttttt}.  
Specifically, a larger timing gap between a cache hit and a miss reduces the bit error rate, and faster attack execution enhances channel capacity.

\subsection{Breaking KASLR}
\label{subsec:breaking_kaslr}

\noindent\textbf{Threat model.}  
We assume an unprivileged attacker who can execute arbitrary code within user space.  
The attacker's goal is to exploit a kernel vulnerability to acquire root privileges.  
To achieve this, the attacker first breaks Kernel Address Space Layout Randomization (KASLR) by determining the precise base address of the kernel image.  

\begin{figure}[t!]
    \centerline{\includegraphics[scale=0.19]{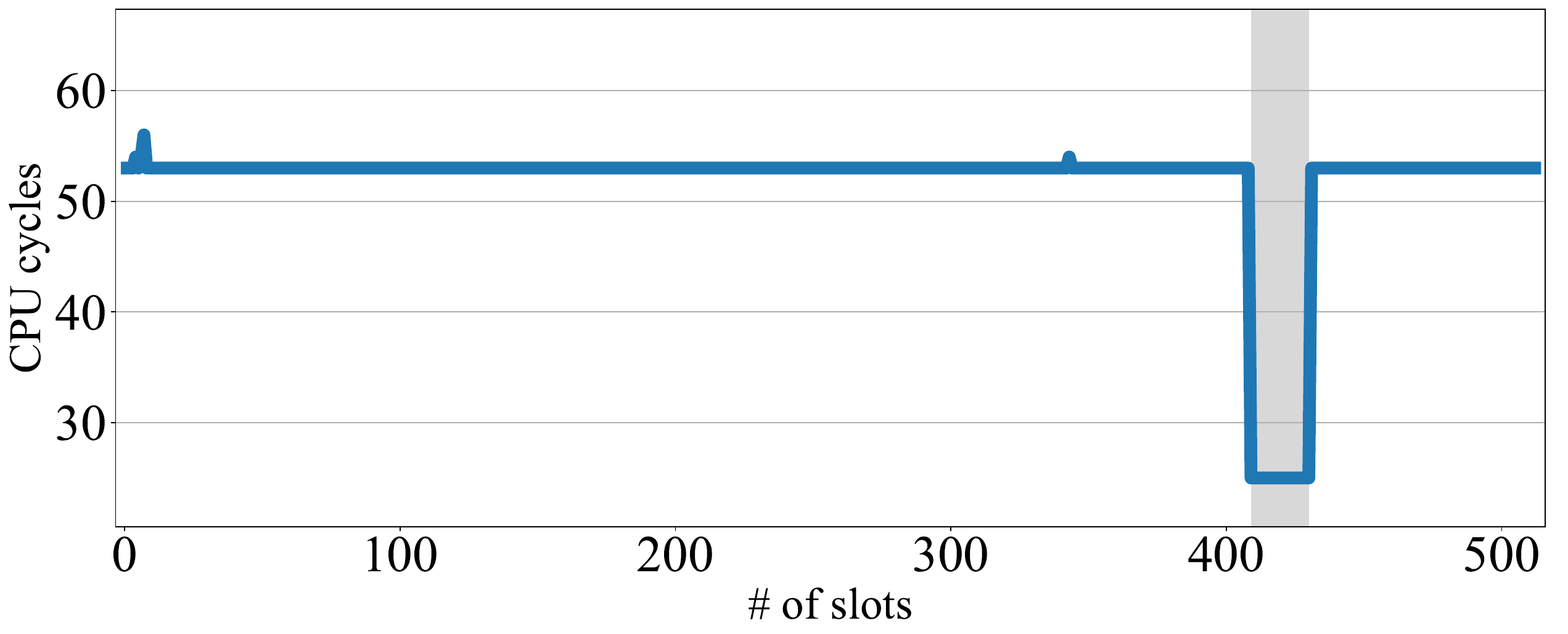}}
    \caption{An experimental results from the \texttt{cldemote}-based KASLR breaking.}
    \label{fig:fig7}
    \vspace{-0.4cm}
\end{figure}

\noindent\textbf{Attack strategy.}  
In modern operating systems, the kernel is loaded into a predetermined memory region rather than a completely random location.  
This region is divided into fixed-size units, known as slots, with each slot representing a potential loading site for the kernel.  
During the boot process, the kernel is randomly placed into one of these slots.  
For instance, in Linux, the kernel is loaded with 2 MiB granularity within the address range \texttt{0xffffffff80000000} to \texttt{0xffffffffc0000000}, resulting in 512 possible slots.  
To identify the slot where the kernel resides, we apply Demote+Time across all slots in this range.  
If the kernel is loaded in a specific slot (that is, if the address is valid), its execution time is noticeably faster than in other slots.  
Conversely, if the kernel is absent from a slot (that is, if the address is invalid), the execution time is slower.  
Figure~\ref{fig:fig7} illustrates the execution latency of Demote+Time for each slot in Linux.  
The gray-shaded section, covering slots 409 through 430, consistently shows low execution times, indicating the kernel is loaded within these slots.  
Consequently, the kernel base address is identified as \texttt{0xffffffffb3200000}.  

\noindent\textbf{Evaluation.}
We evaluate the performance of our attack on Ubuntu 24.04 (kernel version 6.8.0-38).  
Specifically, we performed 1,000 attacks to measure both accuracy and execution time.  
After each set of 100 attacks, we rebooted the system to reset the kernel base address.  
This process resulted in 10 reboots in total.  
During the Demote+Time step, we measured the execution time of 100 consecutive \texttt{cldemote} instructions to reduce potential noise.  
In terms of accuracy, the attack succeeded with a rate of 99.8\% on Ubuntu 24.04, reliably locating the kernel base address.  
For execution time, we computed the average runtime over 1,000 attacks.  
On Ubuntu 24.04, we found the kernel base address in an average of 2.49 milliseconds.

\section{Case Study 2 - Constructing Eviction Set}
\label{sec:case_study2}

In this section, we demonstrate that the \texttt{cldemote} can be utilized not only for performing side-channel attacks but also for constructing eviction sets.
We start by conducting reverse engineering efforts on the directory structure of the Sapphire Rapids processor (Section~\ref{subsec:reversing}).
Based on this knowledge, we propose \texttt{cldemote}-based eviction set construction method on the directory, and evaluate its effectiveness by comparing it with previous works (Section~\ref{subsec:evictionset}).

\subsection{Reverse engineering directory structure}
\label{subsec:reversing}
To carry out \textit{evict}- and \textit{prime}-based attacks on a cross-core scenario where the processor adopts a non-inclusive LLC design, an attacker must target the directory, which maintains inclusivity for the private cache~\cite{yan2019attack, kim2022dprime+}.  
To achieve this, constructing an eviction set $\mathcal{E}$ for a target set $\mathcal{S}$ is crucial.  
With this eviction set, an attacker can reset the states of $\mathcal{S}$ by filling it with each address $\epsilon \in \mathcal{E}$.  
We refer to the addresses in $\mathcal{E}$ as congruent addresses.  

\noindent\textbf{Reverse engineering LLC.}  
Before reverse-engineering the directory structure, understanding the LLC's architectural details, such as set associativity, is essential.  
This is because the directory and LLC share the same number of sets and use an identical address-mapping hash function to locate the LLC or directory slice and set~\cite{yan2019attack}.  
However, the architectural details of Sapphire Rapids remain publicly undisclosed.  

Algorithm~\ref{algorithm:algo2} outlines the reverse-engineering procedure.  
First, we construct an eviction set $\mathcal{E}$ for an LLC set $\mathcal{S}$ that shares the same set and slice bits as the target address \textit{addr} (line 9).  
This can be achieved using previously proposed eviction set construction algorithms~\cite{vila2019theory, song2019dynamically, purnal2021prime+, zhao2024last}.  
Next, we prime $\mathcal{S}$ by accessing each address $\epsilon \in \mathcal{E}$ (lines 1--5).  
Finally, we measure the memory access latency for \textit{addr} using \texttt{rdtsc} (lines 12--15).  
We repeat this procedure while varying the number of congruent addresses \textit{n} in $\mathcal{E}$.  

\begin{figure}[t!]
    \centerline{\includegraphics[scale=0.18]{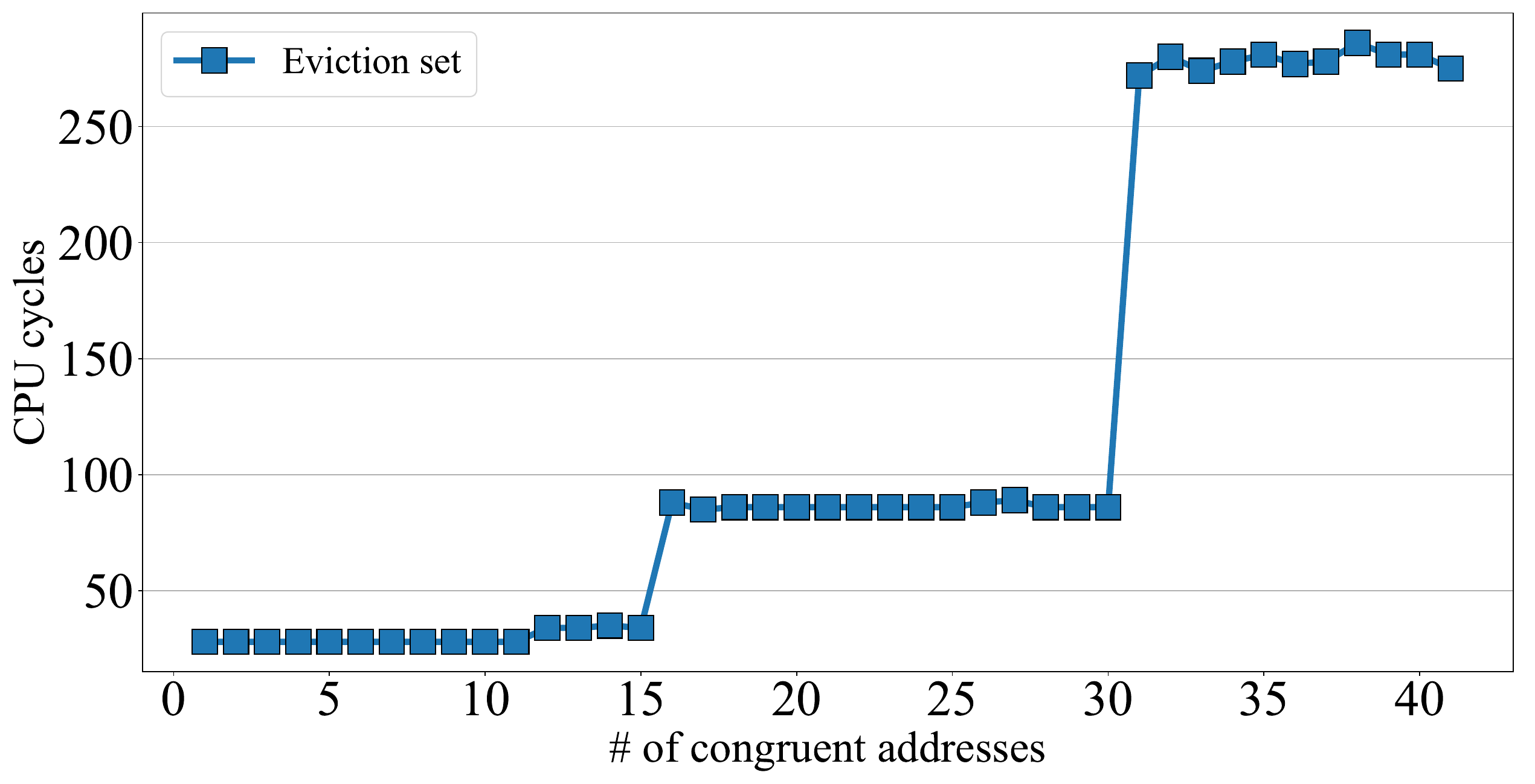}}
    \vspace{-0.3cm}
    \caption{The access latencies from the cache conflict.}
    \label{fig:fig8}
\end{figure}

Figure~\ref{fig:fig8} shows the reverse-engineering results, revealing three distinct latency peaks when the size of $\mathcal{E}$ reaches 12, 16, and 31.  
These latency peaks correspond to an L2 hit, an L3 hit, and a physical memory access, respectively, as illustrated in Figure~\ref{fig:fig_a1}.  
These findings indicate that L1d is 12-way associative, L2 is 16-way associative, and L2 is inclusive of L1d.  
They also confirm that the LLC slice is 15-way associative and non-inclusive of the private caches.  

Additionally, we examine the physical addresses of all congruent addresses to determine the number of sets in each LLC slice.  
Our analysis shows that bits 16:6 of the physical address remain consistent across these congruent addresses.  
This consistency indicates that each LLC slice contains 2048 sets.

\begin{figure}[t!]
    \centerline{\includegraphics[scale=0.19]{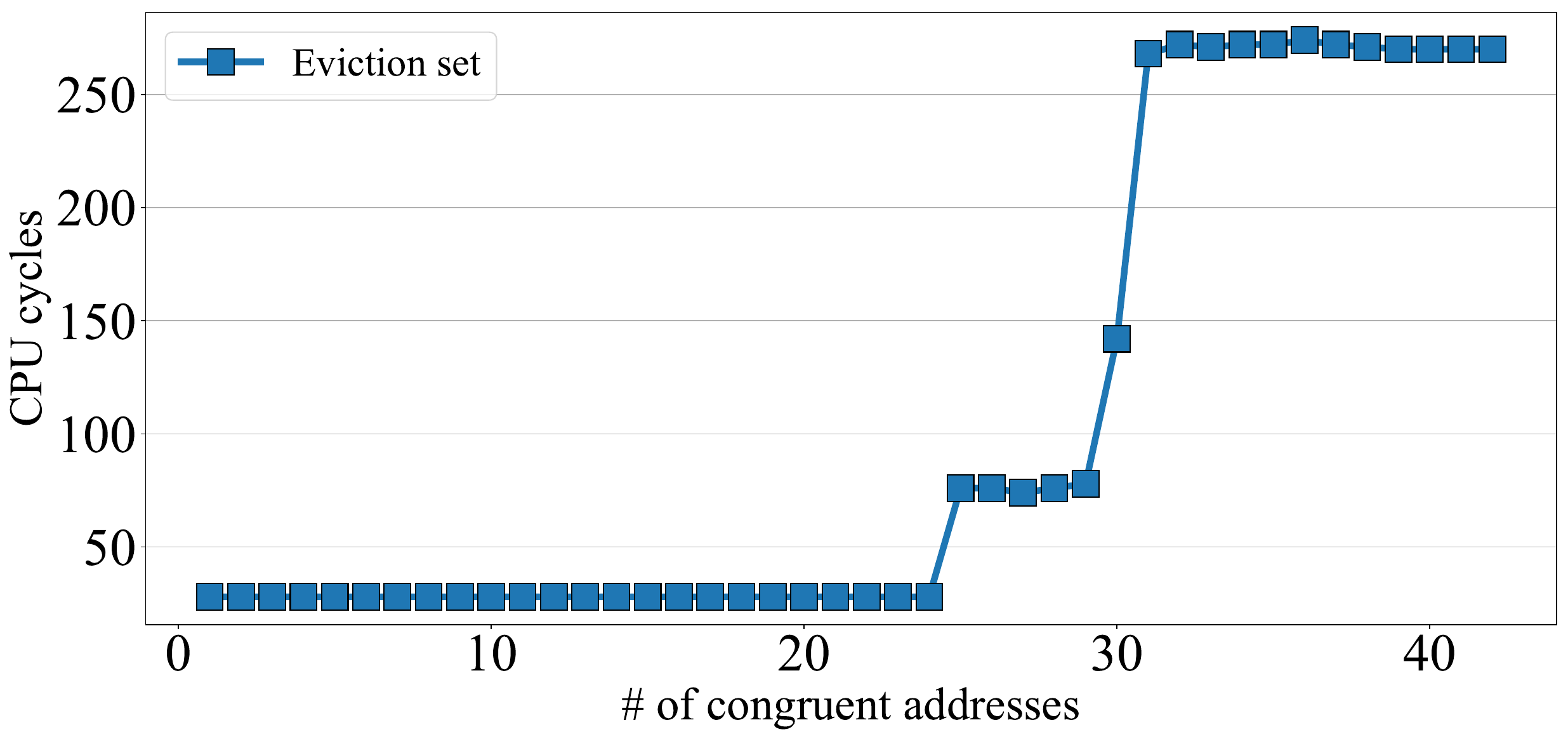}}
    \vspace{-0.2cm}
    \caption{The access latencies from the directory conflicts.}
    \label{fig:fig9}
    \vspace{-0.4cm}
\end{figure}

\noindent\textbf{Comparison to cache parameters from CPUID.}
Cache parameters can be retrieved via the \texttt{CPUID} instruction by querying leaf 0x4 with sub-leaves 0x0, 0x1, 0x2, and 0x3.  
However, our analysis revealed unusual information regarding the LLC.  
While \texttt{CPUID} reports a 30\,MiB LLC with 15-way associativity and 32,768 sets—consistent with Intel's specifications~\cite{Intel2024XeonSilver}—this conflicts with the fact that our processor has 12 physical cores, making 32,768 sets not evenly divisible by 12.  
Our reverse engineering shows that each LLC slice is 15-way associative with 2,048 sets, for a total of 22.5\,MiB of LLC capacity.  
Due to the inaccurate structural information from \texttt{CPUID}, we decided to base our eviction-set construction on these reverse-engineering results, despite the discrepancy with Intel's specifications.

\noindent\textbf{Reverse engineering directory.}
With the knowledge of the LLC structure, we now perform reverse engineering on the directory structure.  
Because the directory shares the same architectural design as the LLC, apart from the number of ways, we seek to discover this parameter.  
We begin our experiment by preparing two threads, referred to as the main thread and the evictor thread.  
We then run them on separate physical cores.  
The main thread first selects a target address $\mathcal{T}$ and accesses it to load a cache line into L1d.  
Next, the evictor thread creates an eviction set $\mathcal{E}$ targeting $\mathcal{T}$ and accesses each address $\epsilon \in \mathcal{E}$ to allocate cache lines into the directory.  
Finally, the main thread measures the access latency for $\mathcal{T}$ using \texttt{rdtsc}.  
If a directory conflict occurs, the main thread experiences increased memory access latency.  
We repeat this procedure, varying the number of congruent addresses in $\mathcal{E}$.  

Figure~\ref{fig:fig9} shows the experimental results.  
The graph reveals two prominent spikes in access latency when the number of congruent addresses reaches 25 and 31.  
These spikes correspond to L3 cache hits and physical memory accesses, respectively (see Figure~\ref{fig:fig_a1}).  
Since a conflict in the directory moves a cache line from the private cache to the shared LLC~\cite{yan2019attack}, this timing behavior indicates that the directory slice has a 25-way associativity.  
Our reverse engineering results are summarized in Table~\ref{table:table8}.

\subsection{Efficient eviction set construction}
\label{subsec:evictionset}
Based on our knowledge of the directory structure, we propose Access+Time to enhance eviction set construction in non-inclusive LLC designs.  

\noindent\textbf{Overcoming the limitations of existing methods.}  
Multiple studies have introduced efficient eviction set construction algorithms to facilitate \textit{evict}- and \textit{prime}-based attacks in various computing environments~\cite{vila2019theory, song2019dynamically, purnal2021prime+, zhao2024last, liu2015last, xue2023ctpp, oren2015spy}.  
However, these approaches primarily focus on inclusive LLCs, because constructing an eviction set for a non-inclusive LLC is more challenging.

\begin{table}[t!]
\centering
\renewcommand{\arraystretch}{1.1}
\caption{Architectural details on Sapphire Rapids.}
\label{table:table8}
\begin{adjustbox}{width=1\columnwidth}
\begin{tabular}{c|c}
\Xhline{3\arrayrulewidth}
          & \textbf{Intel Xeon Silver 4510T}                                \\ \hline\hline
L1 data   & 48KiB, 64 sets, 12 ways                                  \\ 
L1 inst  & 32KiB, 64 sets, 8 ways                                  \\ 
L2        & 2MiB, 2048 sets, 16 ways, inclusive to L1                \\ 
LLC slice & 1.875MiB, 2048 sets, 15 ways, non-inclusive to L1/L2     \\ 
Directory slice  & 3.125MiB, 2048 sets, 25 ways                             \\ \Xhline{3\arrayrulewidth}
\end{tabular}
\end{adjustbox}
\vspace{-0.4cm}
\end{table}

Specifically, processors with a non-inclusive LLC allocate cache lines to the private cache (i.e., L1 and L2) rather than an LLC slice when performing memory accesses.  
Consequently, an attacker must move the cache line from the private cache to the LLC slice to construct an eviction set.  
To address this challenge, prior works~\cite{yan2019attack, zhao2024last, purnal2021prime+} proposed two methods: (1) inducing cache conflicts and (2) using a helper thread.  
However, each method has limitations.  
The first method demands multiple memory accesses to evict a single cache line into the LLC and requires 10 iterations to achieve a low false-negative eviction rate~\cite{yan2019attack}.  
The second method relies on the observation that if two threads on different physical cores access the same data, it resides in the LLC~\cite{yan2019attack, purnal2021prime+}.  
However, this method requires two physical cores, limiting its applicability in single-core attack scenarios.  

To overcome these issues, we propose a new eviction set construction method called Access+Demote, leveraging the \texttt{cldemote} instruction.  
Since \texttt{cldemote} is designed to demote a cache line from the private cache to the shared LLC (the property \textbf{P1}), it can effectively replace the previously proposed approaches.  
Integrating Access+Demote into existing eviction set construction algorithms is straightforward.  
It merely involves replacing prior methods, such as cache conflicts or helper threads, with a single \texttt{cldemote} instruction.

\noindent\textbf{Evaluation.}
To demonstrate the effectiveness of our approach, we compare the helper thread method with Access+Demote in two eviction set construction algorithms,  (1) the widely used baseline algorithm by Vila et al.~\cite{vila2019theory} and  
(2) the Function-as-a-Service oriented algorithm by Zhao et al.~\cite{zhao2024last}.  
For a thorough evaluation, we reproduced both algorithms in our experimental setup using a 4\,KiB page size.  
We then integrated Access+Demote to demote a cache line into the LLC.  
We performed each eviction set construction procedure 100 times.  
We excluded the cache conflict method from our experiment because it is not commonly used, has a high false-negative rate, and requires numerous memory operations.  
Table~\ref{table:table9} presents their success rates and building times.  

The results show that the helper thread achieves success rates exceeding 97\%.  
Similarly, Access+Demote attains comparable or slightly lower success rates.  
However, this method reduces construction time by 36\% through a single \texttt{cldemote} instruction without relying on a helper thread.  
Thus, Access+Demote not only decreases construction time but also overcomes the limitations of requiring two physical cores or multiple memory accesses found in existing methods.

\noindent\textbf{Combining \texttt{cldemote} with Prime+Scope~\cite{purnal2021prime+}.}
While we successfully integrated the Access+Demote method into existing eviction set construction algorithms~\cite{vila2019theory, zhao2024last}, our approach is incompatible with Prime+Scope due to fundamental differences in how Prime+Scope constructs eviction sets.  
Unlike previous works~\cite{song2019dynamically, zhao2024last, liu2015last, xue2023ctpp, oren2015spy}, Prime+Scope begins with an empty set and gradually adds congruent addresses to build an eviction set.  
This process relies on a key observation about the cache's replacement policy:  
memory accesses served from upper-level caches do not influence the replacement policy of the lower-level caches holding the same data.  

To construct an eviction set with Prime+Scope, an attacker first selects a target address $\mathcal{T}$ and accesses it through a helper thread, placing $\mathcal{T}$ in both a private cache and the shared LLC.  
The attacker then accesses candidate addresses and measures the access latency for $\mathcal{T}$ to check if it has been evicted from the cache hierarchy.  
If $\mathcal{T}$ is evicted, the candidate address is deemed congruent.  
Otherwise, the attacker repeats this procedure with other candidates.  

However, if the attacker employs \texttt{cldemote} to move the cache line for $\mathcal{T}$ into the LLC rather than using a helper thread, the private cache does not contain a copy of that cache line.  
Consequently, when the attacker accesses $\mathcal{T}$, a private cache miss occurs, causing a fetch from the LLC and marking the cache line in the shared LLC as more recently used.  
This behavior prevents the attacker from utilizing the replacement policy, making it impossible to construct an eviction set with \texttt{cldemote}.

\begin{table}[t!]
\centering
\renewcommand{\arraystretch}{1.2}
\caption{Performance evaluation of eviction set constructions.}
\label{table:table9}
\begin{adjustbox}{width=1\columnwidth}
\begin{tabular}{cccc}
\Xhline{3\arrayrulewidth}
\multicolumn{2}{c}{}                                                                                                                     & Vila et al.~\cite{vila2019theory} & Zhao et al.~\cite{zhao2024last} \\ \hline\hline
\multicolumn{1}{c|}{\multirow{2}{*}{\begin{tabular}[c]{@{}c@{}}Helper thread\end{tabular}}}       & \multicolumn{1}{c|}{Succ. rate}  & 97\%        & 100\%       \\ \cdashline{2-4}
\multicolumn{1}{c|}{}                                                                                 & \multicolumn{1}{c|}{Build. time} & 82.153 ms   & 19.470 ms   \\ \hline
\multicolumn{1}{c|}{\multirow{2}{*}{\begin{tabular}[c]{@{}c@{}}Access+Demote\end{tabular}}} & \multicolumn{1}{c|}{Succ. rate}  & 97\%        & 98\%        \\ \cdashline{2-4}
\multicolumn{1}{c|}{}                                                                                 & \multicolumn{1}{c|}{Build. time} & 54.842 ms   & 10.792 ms   \\ \Xhline{3\arrayrulewidth}
\end{tabular}
\end{adjustbox}
\vspace{-0.4cm}
\end{table}

\section{Countermeasure}
\label{sec:countermeasure}
We discuss some possible mitigations against our attacks presented in previous sections.

\noindent\textbf{Mitigation against cache attack.}
This attack exploits characteristics of \texttt{cldemote} that we revealed in Section~\ref{sec:key_property}, such as inter-cache state transition, timing difference, and unprivileged access, to infer the victim's memory access patterns.
However, \texttt{cldemote} can only demote a cache line that resides in the same physical core as the attacker (i.e., sibling core).
To mitigate this attack, one possible strategy is to disable SMT.
Disabling SMT prevents threads which have different security domains from being scheduled on the same physical core.
As a result, it becomes impossible to demote cache lines between threads using \texttt{cldemote}, effectively mitigating proposed attacks.

\noindent\textbf{Mitigation against KASLR breaking attack.}
This attack exploits characteristics such as fault suppression and unprivileged access to identify the kernel base address from user space.
Kernel Page-Table Isolation (KPTI), derived from KAISER~\cite{gruss2017kaslr}, is a widely used mitigation technique against KASLR breaking attacks.
The primary strategy of KPTI is to isolate the kernel address space from user space, preventing an unprivileged user from accessing kernel addresses without switching into kernel mode.
However, despite this isolation, certain kernel segments (i.e., the trampoline region) remain mapped in user space, allowing an attacker to infer the kernel base address~\cite{Canella2020KASLR, weber2021osiris, lipp2022prefetch, kim2023avx}.
An alternative mitigation, FLARE~\cite{Canella2020KASLR}, takes a different approach that maps invalid kernel addresses to dummy physical pages.
This strategy ensures all kernel addresses are physically backed, making used kernel memory indistinguishable from unused kernel memory.

\noindent\textbf{Mitigation against \texttt{cldemote}-based attacks.}
Several strategies discussed above offer potential mitigations for specific types of microarchitectural attacks.
However, these strategies do not mitigate all forms of such attacks.
For this purpose, a more fundamental and comprehensive approach is necessary.
One promising solution is to introduce random noise into the execution latency of \texttt{cldemote}.
This random noise disturbs precise timing measurements, making it infeasible for an attacker to infer meaningful data from timing information.
Another solution is to restrict \texttt{cldemote} only to the privileged user rather than allowing unprivileged users to access it.
By doing so, this solution makes performing \texttt{cldemote}-based microarchitectural attacks infeasible.

\section{Systematization of ISA Extensions}
\label{sec:ISA extension driven}
In the previous sections, we analyzed how the \texttt{cldemote} instruction can be exploited to mount microarchitectural attacks.  
However, \texttt{cldemote} is not the only ISA extension that exhibits such vulnerabilities; rather, it serves as a representative example of a broader class of instructions that may share similar properties.  
In this section, we generalize our findings to other ISA extensions, aiming to highlight which common characteristics make them exploitable and how these characteristics combine to enable various attack vectors.  
By providing a taxonomy of exploitable properties and associated attack types, we offer a framework that clarifies the root causes behind both existing and potentially undiscovered vulnerabilities in modern processors.  
This perspective not only underscores the importance of security assessments for \texttt{cldemote}, but also offers guidance for evaluating new or existing ISA extensions in future designs.

We examine how ISA extensions can be exploited for microarchitectural attacks.
Our analysis identifies five key characteristics that make these extensions exploitable (Section~\ref{subsec:characteristic}).
We then categorize four types of attacks based on their combinations (Section~\ref{subsec:systematic}).
The results of our analysis are summarized in Table~\ref{table:table22}.

\subsection{Characteristics of exploitable ISA extensions}
\label{subsec:characteristic}

Prior works have shown that ISA extensions can be exploited for microarchitectural attacks.  
For instance, extensions such as \texttt{Intel TSX}~\cite{jang2016breaking, kim2022dprime+, disselkoen2017prime+}, \texttt{umonitor/umwait}~\cite{zhang2023m}, \texttt{vmaskmovd}~\cite{choi2023avx,kim2023avx}, \texttt{prefetch}~\cite{gruss2016prefetch, lipp2022prefetch, guo2022adversarial}, \texttt{clflush}~\cite{gruss2016flush+, yarom2014flush+}, and \texttt{movnt}~\cite{weber2021osiris} have been used to extract security-sensitive data.  
To understand how these extensions enable such attacks, we analyze each extension and its specific role in the attacks.  

Our analysis identifies five key characteristics that make ISA extensions exploitable, as follows.  

\noindent\textbf{Characteristic 1: Unprivileged access} ($\mathcal{U}$).  
This means the extension can be used by unprivileged users.  
If $\mathcal{U}$ is not satisfied, an ISA extensions-driven attack cannot be performed at the unprivileged level.  
As a result, all ISA extensions exploited in side-channel attacks exhibit this characteristic.  

\noindent\textbf{Characteristic 2: Inter-cache state transition} ($\mathcal{I}$).  
This refers to the ability to induce cache state transitions between private caches and the LLC.  
Among all ISA extensions, only \texttt{cldemote} exhibits this behavior by moving a cache line from the private cache to the LLC.  
Since \texttt{cldemote} is a memory operation, its execution implicitly triggers address translation, enabling potential changes to TLB state.  

\noindent\textbf{Characteristic 3: Memory-cache state transition} ($\mathcal{M}$).  
This indicates bidirectional cache transitions between physical memory and the cache.  
In the memory-to-cache direction, extensions like \texttt{prefetch} and \texttt{vmaskmovd} load data from physical memory into the cache.  
In the cache-to-memory direction, extensions such as \texttt{clflush} and \texttt{movnt} flush a cache line, updating the corresponding data in memory.  
These operations can also alter the TLB state because they involve address translation.  

\noindent\textbf{Characteristic 4: Detecting cache state transition} ($\mathcal{D}$).  
This allows an attacker to identify cache state transitions without relying on timing, power, or thermal measurements.  
For instance, \texttt{umonitor/umwait} detects changes in a monitored cache line by observing the carry flag, and \texttt{XBEGIN/XEND} reveals whether a cache line has been evicted from the L1 or LLC by examining its abort status.  

\noindent\textbf{Characteristic 5: Fault suppression} ($\mathcal{S}$).  
This means the ISA extension suppresses faults that would normally occur when accessing inaccessible memory addresses.  
In particular, extensions such as \texttt{XBEGIN/XEND}, \texttt{prefetch}, \texttt{vmaskmovd}, and \texttt{cldemote} exhibit this behavior, effectively suppressing faults in the attack process.

\begin{table*}[t!]
\centering
\renewcommand{\arraystretch}{1.0}
\caption{Characteristics of exploitable ISA extensions and four types of attacks resulting from their combinations.}
\label{table:table22}
\begin{adjustbox}{width=0.99\textwidth}
\begin{tabular}{c:ccccc:cccc}
\Xhline{3\arrayrulewidth}
\multirow{3}{*}{\textbf{Instruction}} & \multicolumn{5}{c:}{\textbf{Characteristic}}                                                                                                                                                                                                                                                                                                                         & \multicolumn{4}{c}{\textbf{Microarchitectural attack}}                  \\ 
                             & \begin{tabular}[c]{@{}c@{}}\textbf{Unprivileged}\\ \textbf{access}\end{tabular} & \begin{tabular}[c]{@{}c@{}}\textbf{Inter-cache}\\ \textbf{state transition}\end{tabular} & \begin{tabular}[c]{@{}c@{}}\textbf{Cache-memory}\\ \textbf{state transition}\end{tabular} & \begin{tabular}[c]{@{}c@{}}\textbf{Detecting cache}\\ \textbf{state transition}\end{tabular} & \begin{tabular}[c]{@{}c@{}}\textbf{Fault}\\ \textbf{suppression}\end{tabular} & \begin{tabular}[c]{@{}c@{}}\textbf{Cache}\\ \textbf{attack}\end{tabular} & \begin{tabular}[c]{@{}c@{}}\textbf{Noise-free}\\ \textbf{cache attack}\end{tabular} & \begin{tabular}[c]{@{}c@{}}\textbf{Faultless}\\ \textbf{KASLR breaking}\end{tabular} & \begin{tabular}[c]{@{}c@{}}\textbf{Fast eviction}\\ \textbf{set construction}\end{tabular} \\
                             & ($\mathcal{U}$)                                                           & ($\mathcal{I}$)                                                                    & ($\mathcal{M}$)                                                                     & ($\mathcal{D}$)                                                                        & ($\mathcal{S}$)                                                         & ($\mathcal{U}$ $\land$ ($\mathcal{I}$ $\lor$ $\mathcal{M}$))                                            & ($\mathcal{U}$ $\land$ $\mathcal{D}$)                                                    & ($\mathcal{U}$ $\land$ ($\mathcal{I}$ $\lor$ $\mathcal{M}$) $\land$ $\mathcal{S}$ )                                                 & ($\mathcal{U}$ $\land$ $\mathcal{I}$)                                                                   \\ \hline\hline
\texttt{XBEGIN/XEND}                  & \CIRCLE    & \Circle      &    \Circle   &    \CIRCLE     &    \CIRCLE        &  \Circle        &       \CIRCLE       &     $\,\;$\CIRCLE$^{\dagger}$      &  \Circle      \\
\texttt{umonitor/umwait}              & \CIRCLE    & \Circle      &    \Circle   &    \CIRCLE     &    \Circle        &   \Circle       &       \CIRCLE       &     \Circle      &  \Circle       \\
\texttt{vmaskmovd}                    & \CIRCLE    & \Circle      &    \CIRCLE   &    \Circle     &    \CIRCLE        &   \RIGHTcircle  &       \Circle       &     \CIRCLE      &  \Circle       \\
\texttt{prefetch}                     & \CIRCLE    & \Circle      &     \CIRCLE  &    \Circle     &    \CIRCLE        &   \CIRCLE       &       \Circle       &     \CIRCLE      &  \Circle      \\
\texttt{clflush}                      & \CIRCLE    & \Circle      &     \CIRCLE  &    \Circle     &    \Circle        &   \CIRCLE       &       \Circle       &     $\,\;$\RIGHTcircle$^{\dagger}$ &  \Circle        \\
\texttt{movnt}                        & \CIRCLE    & \Circle      &     \CIRCLE  &    \Circle     &    \Circle        &   \CIRCLE       &       \Circle       &     $\,\;$\RIGHTcircle$^{\dagger}$ &  \Circle        \\
\texttt{cldemote}                     & \CIRCLE    &  \CIRCLE     &     \Circle  &    \Circle     &    \CIRCLE        &   \CIRCLE       &       \Circle       &     \CIRCLE      &  \CIRCLE       \\ \Xhline{3\arrayrulewidth}
\end{tabular}
\end{adjustbox}
\footnotesize{\begin{flushleft} Symbols (\CIRCLE$\;$ or \Circle) indicate whether the ISA extension meets the specified characteristic. The symbol (\RIGHTcircle) indicates that the attack is feasible, but has not yet been proposed. The symbol ($^{\dagger}$) indicates that additional instructions are required to perform the attack successfully.
\end{flushleft} }
\vspace{-0.5cm}
\end{table*}

\subsection{Categorization of possible attacks}
\label{subsec:systematic}
We categorize four types of attacks based on the combinations of the characteristics identified in Section~\ref{subsec:characteristic}.

\noindent\textbf{Cache attack}.
This attack exploits differences in the measurements of the system state—such as timing~\cite{kurth2020netcat, shin2020inferring}, power consumption~\cite{lipp2021platypus, wang2022hertzbleed, wang2023dvfs}, or thermal data~\cite{kim2022thermalbleed}—observed during memory access to infer the victim's behavior.  
It typically follows a Reset-Wait-Measurement sequence:  
1) Reset.  
An attacker initializes the cache to a known state.  
2) Wait.  
An attacker waits for the victim's memory access to occur.  
3) Measurement.  
An attacker measures the state to infer changes in the cache state.  
Because the reset and measurement steps require changing the cache state, ISA extensions that trigger cache state transitions (either $\mathcal{I}$ or $\mathcal{M}$) can be used in these steps.  
In addition, cache attacks are usually performed at the unprivileged level ($\mathcal{U}$).  
Therefore, ISA extensions satisfying $\mathcal{U}$ $\land$ ($\mathcal{I}$ $\lor$ $\mathcal{M}$) can be used to conduct a cache attack.  

\noindent\textbf{Noise-free cache attack}.
Unlike traditional cache attacks, this approach exploits architectural signals, such as the carry flag or TSX abort status, to function reliably in noisy environments.  
It follows the same Reset-Wait-Measurement pattern as a conventional cache attack but differs slightly in implementation.  
During the reset step, the attacker begins monitoring a specific cache line or set.  
The attacker then waits for an architectural event that occurs right after the monitored cache line or set undergoes a state transition.  
By detecting this event, the attacker can accurately infer the victim's cache behavior without relying on iterative measurements at fixed intervals.  
However, this attack depends on instructions like \texttt{umonitor/umwait} or \texttt{XBEGIN/XEND} to detect cache state transitions ($\mathcal{D}$).  
It also typically operates at the unprivileged level ($\mathcal{U}$).  
Hence, ISA extensions satisfying both $\mathcal{U}$ and $\mathcal{D}$ ($\mathcal{U}$ $\land$ $\mathcal{D}$) can be used to launch noise-free cache attacks.  

\noindent\textbf{Faultless KASLR breaking attack}.
KASLR breaking exploits TLB properties to determine the kernel image base address.  
Specifically, modern processors allow an unprivileged attacker to allocate a TLB entry for a valid kernel address.  
By measuring memory access latency, the attacker can discover whether a TLB entry exists for that kernel address.  
ISA extensions that induce TLB state transitions (i.e., $\mathcal{I}$ or $\mathcal{M}$) can be used to exploit this behavior.  
However, user-space accesses to kernel memory typically cause page faults.  
Hence, ISA extensions that suppress faults ($\mathcal{S}$) facilitate faultless attacks with higher accuracy and less overhead.  
Therefore, ISA extensions satisfying $\mathcal{U}$ $\land$ ($\mathcal{I}$ $\lor$ $\mathcal{M}$) $\land$ $\mathcal{S}$ can be leveraged for faultless KASLR breaking.  
Although ISA extensions do not inherently suppress all faults, attackers can still achieve faultless attacks by using techniques like Intel TSX or speculative execution.  
For instance, instructions such as \texttt{clflush} or \texttt{movnt} cannot suppress faults on their own, but can be combined with external fault suppression methods to achieve faultless attacks.  

\noindent\textbf{Fast eviction set construction}.
Constructing eviction sets in a non-inclusive LLC is not straightforward because attackers must be able to move cache lines from private caches to the LLC.  
Prior work has employed helper threads or induced cache conflicts, but these methods either need multiple physical cores or require many memory operations, prolonging eviction set building time.  
ISA extensions that satisfy inter-cache state transition ($\mathcal{I}$) enable faster eviction set construction in non-inclusive LLCs, effectively overcoming these previous challenges.  
For instance, \texttt{cldemote} quickly demotes a cache line to the LLC, significantly speeding up eviction set creation.  
Hence, ISA extensions that meet $\mathcal{U}$ $\land$ $\mathcal{I}$ can be used to construct eviction sets more efficiently.

\section{Related Work}
\label{sec:Related work}
We discuss various side-channel attacks that exploit ISA extensions.

\noindent\textbf{Memory sharing-based cache side-channel attack.}
This attack exploits a shared cache line between an attacker and a victim to infer the victim's secret-dependent memory access patterns.
To achieve this, various ISA extensions have been employed to reset the state of the shared cache line.
Traditionally, \texttt{clflush}~\cite{pereida2016make, gruss2016flush+, yarom2014flush+} has served this purpose.
However, other extensions can achieve the similar effect.

Weber et al.\cite{weber2021osiris} demonstrated that \texttt{movntdq} modifies the shared cache state in both same-core and cross-core attack scenarios.
However, its functionality is limited to writeable pages, restricting its applicability to covert channel attacks.
Guo et al.\cite{guo2022adversarial} showed that \texttt{prefetchw} changes the cache state regardless of page permissions.
However, it is exploitable in cross-core attack scenarios with Cascade Lake processors and earlier generations.
In contrast, our work exploits \texttt{cldemote}, which is available from Sapphire Rapids processors and work in the same-core attack scenarios.

Similar to our work, Rauscher et al.~\cite{Rauscher2025systematic} explored \texttt{cldemote}-based cache attacks, focusing on systematic evaluation with existing methods.
However, our work identified exploitable characteristics of \texttt{cldemote}, and proposed side-channel attacks targeting both the cache and TLB.
We also reverse-engineered the LLC and directory structure of the Sapphire Rapids processor to validate the Access+Demote method in constructing eviction sets on a non-inclusive LLC.
Finally, we analyzed ISA extensions, identified five exploitable characteristics, and categorized four distinct attack types based on their combinations.

\noindent\textbf{KASLR breaking attack.}
This attack aims to infer the kernel base address by exploiting microarchitectural components like the TLB or branch target buffer.
However, accessing a kernel address from user space triggers a page fault that must be handled.
Prior works addressed this fault with ISA extensions.
For instance, Gruss et al.\cite{gruss2016prefetch} used software-based prefetch instructions, Jang et al.\cite{jang2016breaking} leveraged Intel TSX, and Choi et al.\cite{choi2023avx} utilized AVX extensions to bypass KASLR.

In addition, several works relied on speculative execution to suppress fault.
Evtyushkin et al.\cite{evtyushkin2016jump} and Maisuradze et al.\cite{maisuradze2018speculose} exploited branch prediction.
More recently, Phantom\cite{wikner2023phantom} and Inception~\cite{trujillo2023inception} exploit speculatively executed instructions at much earlier stages of pipeline (i.e., before the instruction decode stage).
Contrary, meltdown-type attacks, including Meltdown~\cite{lipp2018meltdown}, Data Bounce~\cite{schwarz2019store}, EchoLoad~\cite{cassell2017nessie}, and FlushConflict~\cite{weber2021osiris}, use fault handlers or Intel TSX to achieve similar goals.
Our work exploits Intel's new \texttt{cldemote}, an ISA extension with fault suppression capability.
Unlike prior works, we target the TLB for store operations, which was first introduced since the Sapphire Rapids processor.

\noindent\textbf{Eviction set construction.}
To perform the \textit{evict}- and \textit{prime}-based attacks, constructing eviction set is crucial.
Early research primarily targeted Intel processors featuring an inclusive LLC design.
Liu et al.\cite{liu2015last} and Irazoqui et al.\cite{irazoqui2015s} used huge pages to create eviction sets without knowledge of physical address.
Other approaches, such as group testing~\cite{vila2019theory, qureshi2019new, song2019dynamically}, grouped candidate addresses and tested whether it is irrelevant ones.
Recently, with the transition to non-inclusive LLC design, eviction set construction techniques have been complicated.
Yan et al.~\cite{yan2019attack} introduced a novel method on the directory (i.e., it has inclusivity for private caches) by using helper threads and inducing cache conflicts.
Additionally, several studies have reverse-engineered hash functions for cache slices\cite{yarom2015mapping, irazoqui2015systematic, maurice2015reverse} and TLB sets~\cite{Gras2018TranslationAttacks} to directly map a cache line to a target set.
In this work, we present a novel eviction set construction method called Access+Demote, designed for non-inclusive LLCs.
This method not only simplifies the eviction set construction procedure but also reduces the construction time by 36\%.

\noindent\textbf{Side-channel attacks with other ISA extensions.}
Zhang et al.\cite{zhang2023m} leveraged the \texttt{umwait} and \texttt{tpause} instructions, which place processor cores in low-power states, to observe targeted cache lines via the carry flag, conducting noise-free cache attacks.
Similarly, Rauscher et al.~\cite{rauscher2024idleleak} leveraged the \texttt{tpause} instruction to observe the side effect of the idle state, enabling keystroke timing attack and the websites/videos fingerprinting.
In addition, Rauscher et al.~\cite{rauscher2024cross} proposed an Inter-Processor Interrupt (IPI) side-channel by using user interrupt and IPI virtualization extensions to detect system interrupts.
Based on these observations, they performed covert channel attack, keystroke timing attack, and website fingerprinting attack.

\section{Conclusion}
\label{sec:Conclusion}
In this paper, we explored Intel's new \texttt{cldemote} extension, which is designed to move a cache line from private caches to the LLC.
We identified several characteristics that make \texttt{cldemote} exploitable in microarchitectural attacks, and demonstrated its effectiveness through two case studies: side-channel attacks and eviction set construction.
First, we proposed Flush+Demote, which constructs a covert channel with a bandwidth of 2.84 Mbps, and Demote+Time, which breaks KASLR in 2.49 ms on Linux.
We then showed that leveraging \texttt{cldemote} for eviction set construction procedure reduces building time by 36\%, addressing the challenges in a non-inclusive LLC design.
We also identified five characteristics that make ISA extensions exploitable and categorize four possible types of attacks based to their combinations.
This finding shows that introducing a new ISA extension without security evaluation can cause security threats.

\section*{Ethics Considerations}
We used publicly available datasets with no personally identifiable information, and we adhered to relevant ethical guidelines throughout our research. 
Our attacks discovered during the study were reported to Intel on January 23, 2025 under a responsible disclosure process. 
Overall, we conducted our work with responsibility and transparency.

\section*{Open Science}
In line with the conference’s policy, we will share our code and data through a public repository upon publication. 
The repository will include source code, libraries, and instructions for reproducing our experiments. 
We will also document data collection and processing steps, alongside guidelines for reuse.




\bibliographystyle{plain}
\bibliography{references.bib}


\section*{Appendix A. Checking the support for CLDEMOTE instruction}
\label{appendix:support}
Intel processors provide a way to check if they support \texttt{cldemote} through the \texttt{CPUID} instruction.
The 25th bit of the ECX register in the return value of \texttt{CPUID}, where the leaf\footnote{Leaf refers to EAX register.} is 0x7 and sub-leaf\footnote{Sub-leaf refers to ECX register.} is 0x0, indicates the availability of the \texttt{cldemote} instruction.
That is, if bit 25 is 0, the processor does not support the \texttt{cldemote} instruction, and it will function as a \texttt{nop}. 
Conversely, if bit 25 is 1, the processor supports the \texttt{cldemote} instruction.

\section*{Appendix B. Distinguishing page table level}
\label{appendix:ptl}

We show how Demote+Time can reveal the page table level at which address translation aborts.
Several previous works have already demonstrated that software-based prefetch instructions~\cite{gruss2016prefetch, lipp2022prefetch} can distinguish page table levels.
However, Demote+Time achieves this by leveraging the newly introduced \texttt{cldemote} instruction on the latest 5th generation of Intel Xeon processor.
It is noteworthy that from the 5th Gen Intel Xeon processor, previous techniques~\cite{gruss2016prefetch, lipp2022prefetch} are unable to leak this information anymore.

To demonstrate the feasibility of Demote+Time in distinguishing page table levels, we design and conduct an experiment.
In this experiment, we prepare 5 different virtual addresses from an inaccessible address space, where each address ends up on different page table levels (i.e., PGD, P4D, PUD, PMD, and PT).
Specifically, these addresses are configured with the Present bit set to 0 and the User/Supervisor bit set to 0.
For each address, we measure the execution time of the \texttt{cldemote} instruction 1 million times using the \texttt{rdtsc}.

On the Intel processor, the TLB caches the address translation only when the virtual address being accessed has the Present bit set to 1 and the Reserved bits set to 0~\cite{Intel2023SDMv3}.
Therefore, we do not need to perform the TLB flush as memory accesses to these addresses consistently trigger a page table walk.

\begin{figure}[t!]
    \centerline{\includegraphics[width=\columnwidth, height=3.8cm]{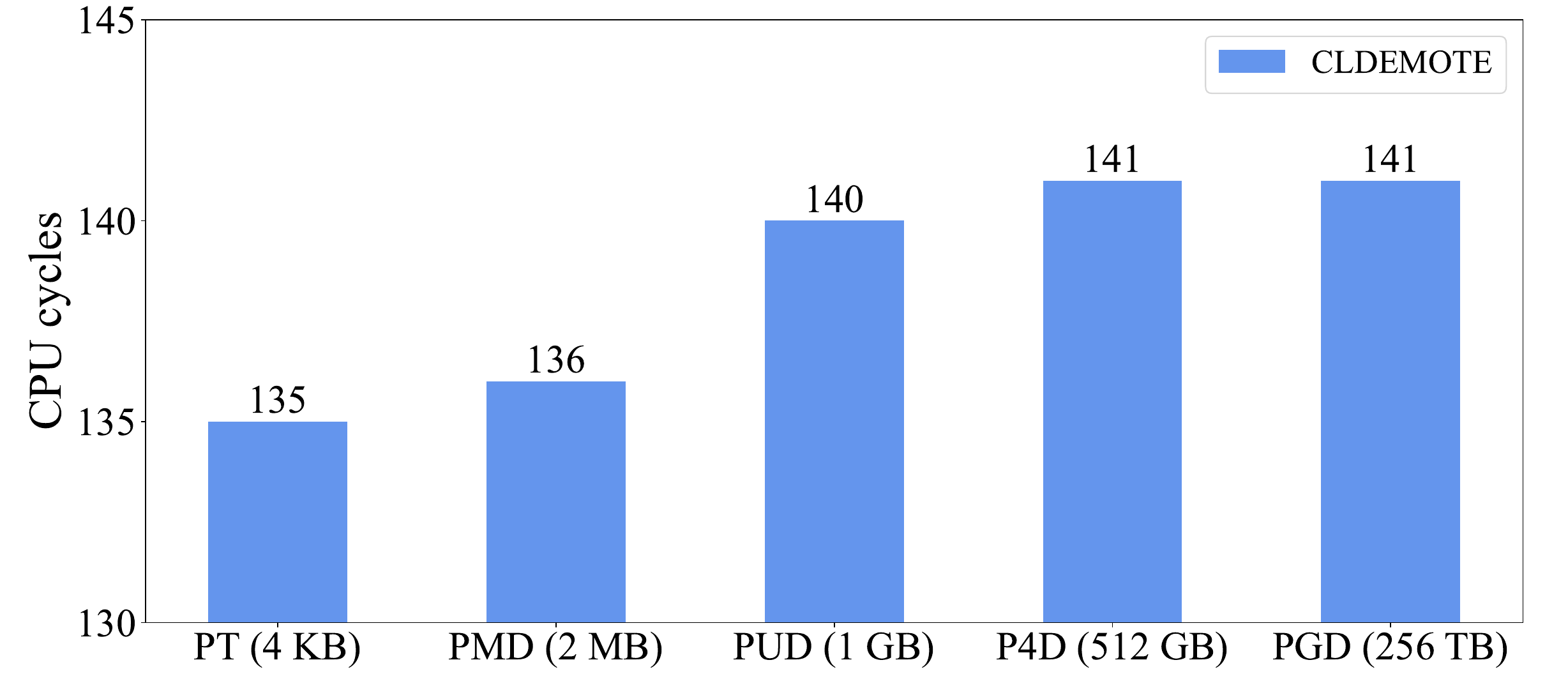}}
    \vspace{-0.2cm}
    \caption{Latency of \texttt{cldemote} across page table levels.}
    \label{fig:fig3}
    \vspace{-0.2cm}
\end{figure}

\begin{figure}[t!]
    \centerline{\includegraphics[width=\columnwidth, height=3.8cm]{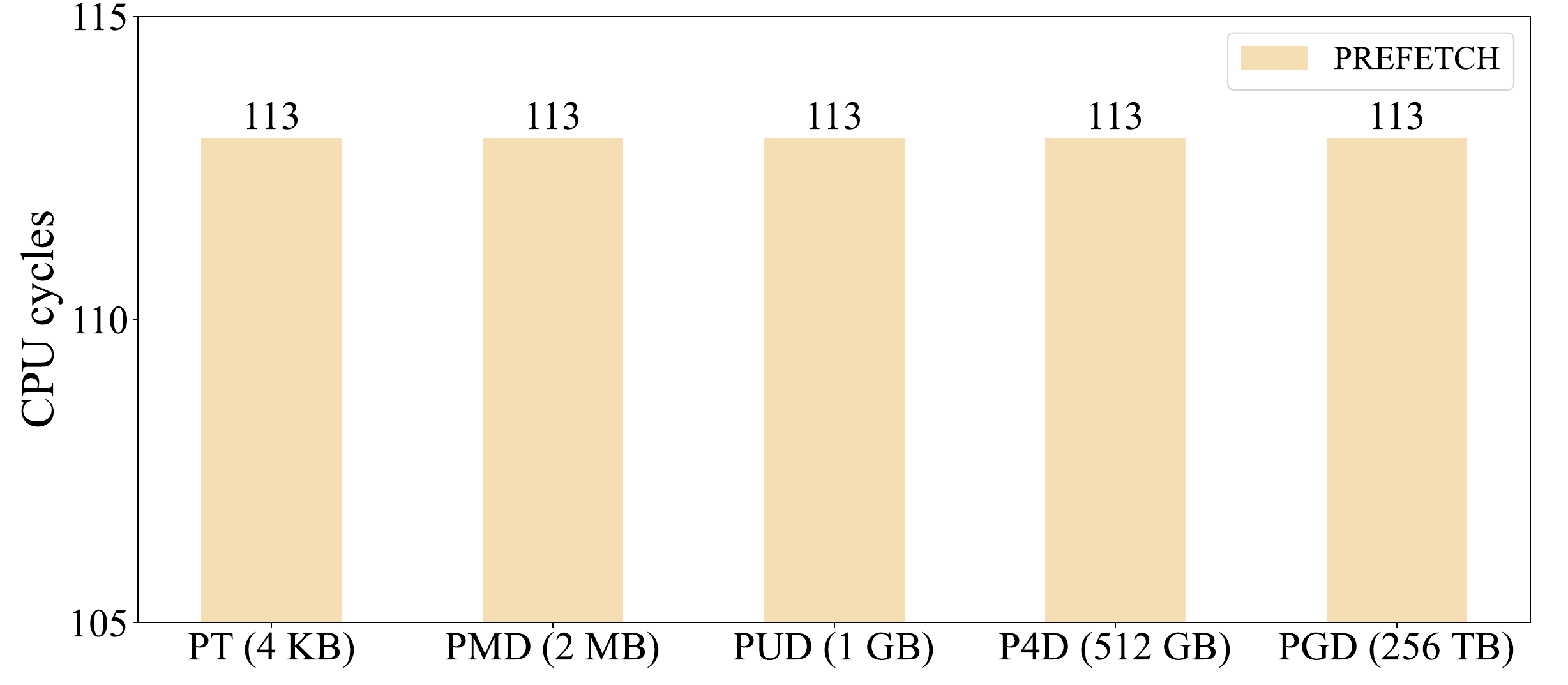}}
    \vspace{-0.2cm}
    \caption{Latency of \texttt{prefetcht2} across page table levels.}
    \label{fig:fig4}
    \vspace{-0.4cm}
\end{figure}

\smallskip\noindent\textbf{Evaluation.}
Figure~\ref{fig:fig3} illustrates the experimental results for the five different virtual addresses, displaying the average execution latency of \texttt{cldemote} at the top of the bar plot.
The results indicate that as the number of page table mappings for an address increases, the time required for address translation through the page table walk decreases.
This observation is consistent with the findings of Lipp et al.~\cite{lipp2022prefetch}, where they observed similar behavior with software-based prefetch instruction at various page table levels. 

Although the timing information may seem counterintuitive when considering the number of page tables mapped to an address, this result can be explained by the behavior of the Paging-Structure Caches~\cite{barr2010translation}. 
According to Intel's documentation~\cite{Intel2023SDMv3}, the paging-structure caches initially attempt to retrieve the physical address for the Page Table (PT) by indexing with virtual address bits 56:21.
It then proceeds to higher levels of the cache hierarchy, up to the Page Global Directory (PGD) by indexing with virtual address bits 56:48.
Consequently, the execution time of \texttt{cldemote} is measured as the shortest, at 113 cycles, when address translation is completed at the PT.

When we repeat this experiment with software-based prefetch instructions like \texttt{prefetcht0/1/2}, \texttt{prefetchnta}, and \texttt{prefetchw}, we cannot observe the same results as with \texttt{cldemote}.
Figure~\ref{fig:fig4} shows an experimental result obtained with \texttt{prefetcht2}.
This experimental result indicates that unlike with \texttt{cldemote}, software-based prefetch instructions do not allow us to identify the specific page table level at which address translation is aborted.

As a result, the execution time of \texttt{prefetcht2} is consistently measured across different page table levels
This is because $dTLB_{load}$ allocates a TLB entry even when the address translation fails to retrieve a physical address, as the address is not backed by physical memory (cf. Section~\ref{subsec:demote+time}).
As a result, an attacker cannot infer the page table levels using $dTLB_{load}$ because the TLB consistently produces TLB hits without triggering a page table walk.

\begin{figure}[t!]
    \centerline{\includegraphics[scale=0.19]{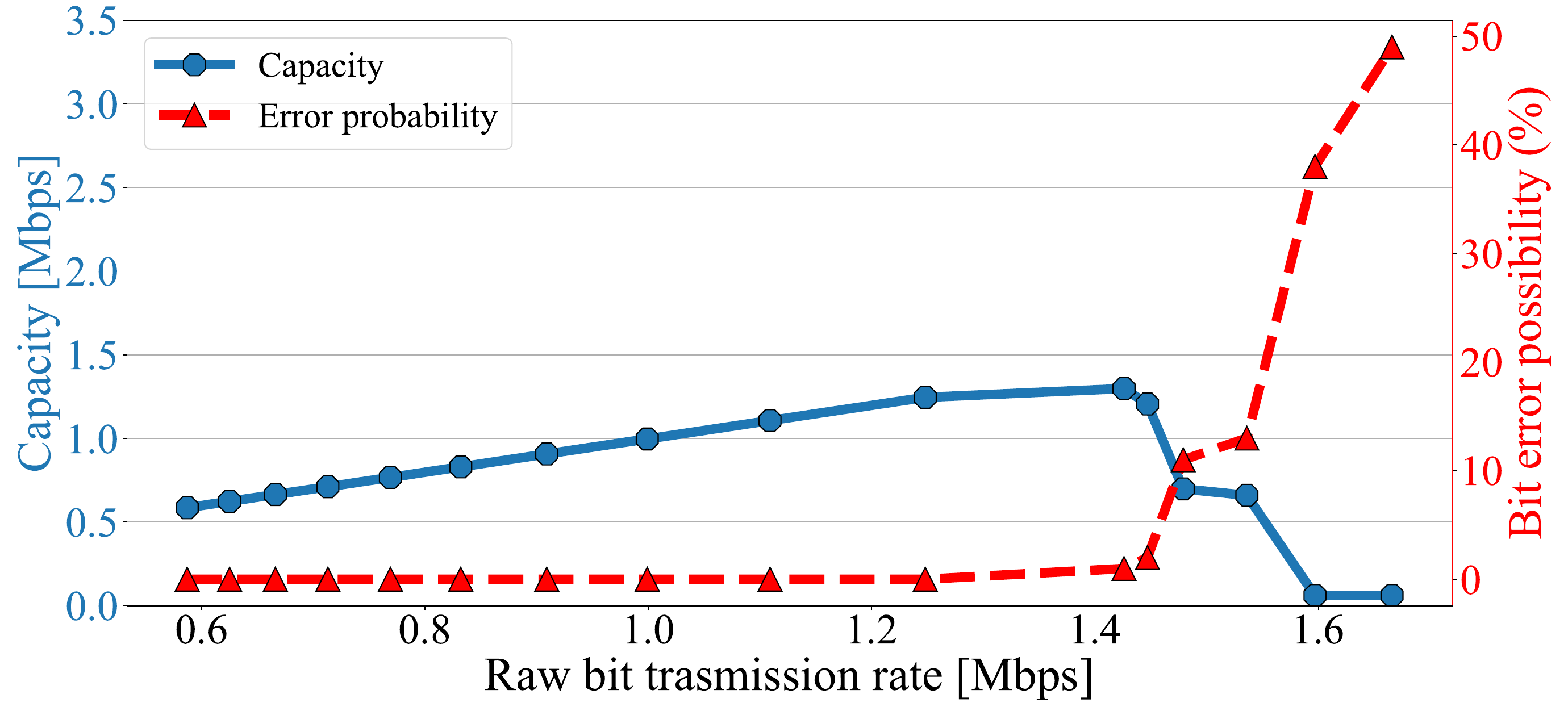}}
    \vspace{-0.2cm}
    \caption{The channel capacity and bit error possibility for Flush+Reload covert channel.}
    \label{fig:fig_a_cc2}
    \vspace{-0.4cm}
\end{figure}

\begin{figure}[t!]
    \centerline{\includegraphics[scale=0.19]{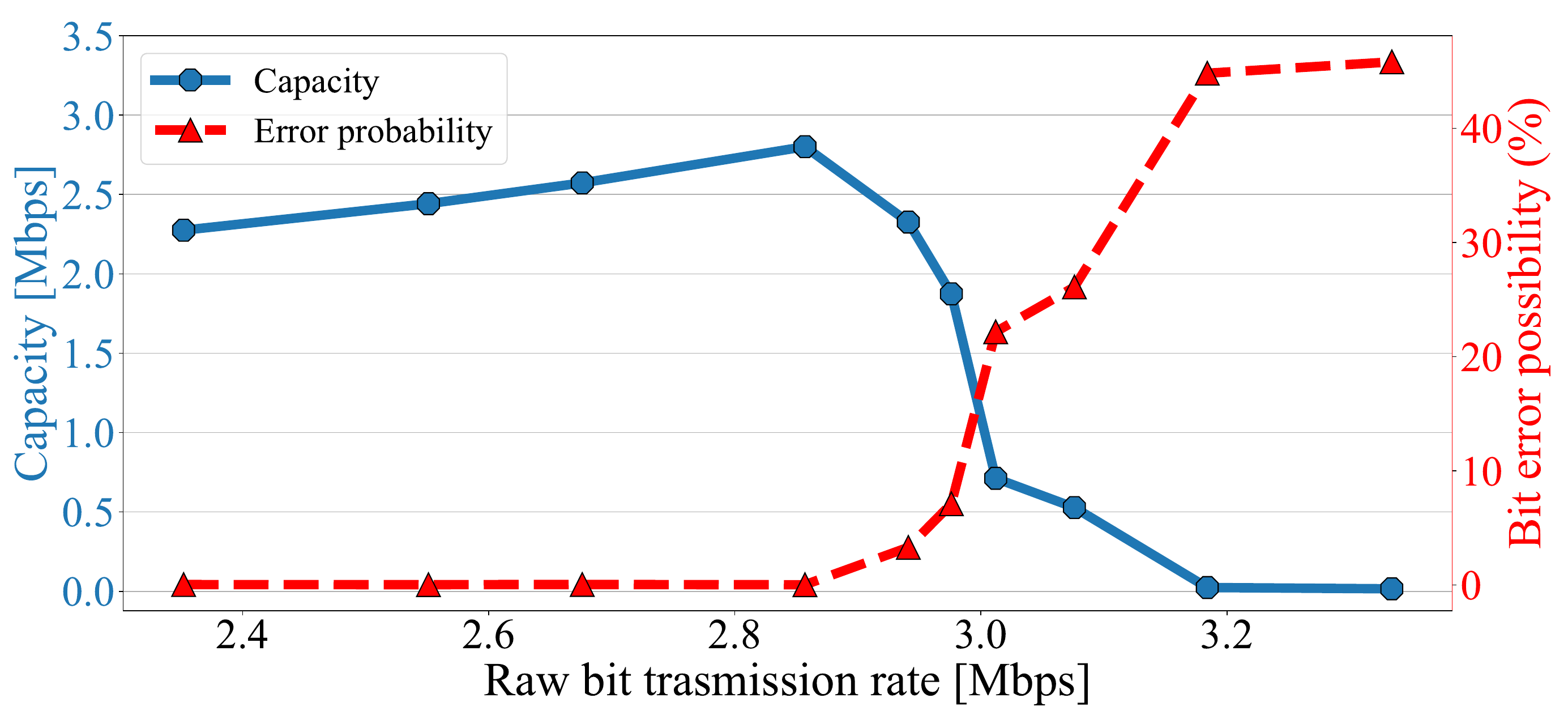}}
    \vspace{-0.2cm}
    \caption{The channel capacity and bit error possibility for Flush+Flush covert channel.}
    \label{fig:fig_a_cc1}
    \vspace{-0.4cm}
\end{figure}

\begin{algorithm}[t!]
\DontPrintSemicolon
  \SetKwFunction{Fattack}{\textit{Prime}}
  
  \KwInput{Target address \textit{addr}, the maximum number of congruent addresses  \textit{N}}
  \KwOutput{The measured timing information \textit{result}}

  \SetKwProg{Fn}{procedure}{:}{end}
  \Fn{\Fattack{$\mathcal{E}$}}{
          
      \For{ \textit{each} $\epsilon$ \textbf{\textit{in}} $\mathcal{E}$} 
        {
           \texttt{maccess}($\epsilon$)
        }
  }

  \texttt{\\}

  \For{\textit{n} $\leftarrow1$ \KwTo \textit{N}}
    {
        tmp $\leftarrow0$ 
       
       $\mathcal{E}$ $\leftarrow$ \texttt{construct\_EVSet}(\textit{addr}, \textit{n})
              
       \For{\textit{i} $\leftarrow1$ \KwTo $100\;000$}
            {
                       
                \texttt{Prime}($\mathcal{E}$)
            
                start $\leftarrow$ \textit{rdtsc()}

                \texttt{maccess}(\textit{addr})
                
                end $\leftarrow$ \textit{rdtsc()} - start

                tmp $\leftarrow$ tmp + end
            }
                
            result[\textit{n}] $\leftarrow$ tmp
    }

\caption{Reverse engineering procedure for LLC.}
\label{algorithm:algo2}
\end{algorithm}

\begin{figure}[t!]
    \centerline{\includegraphics[scale=0.19]{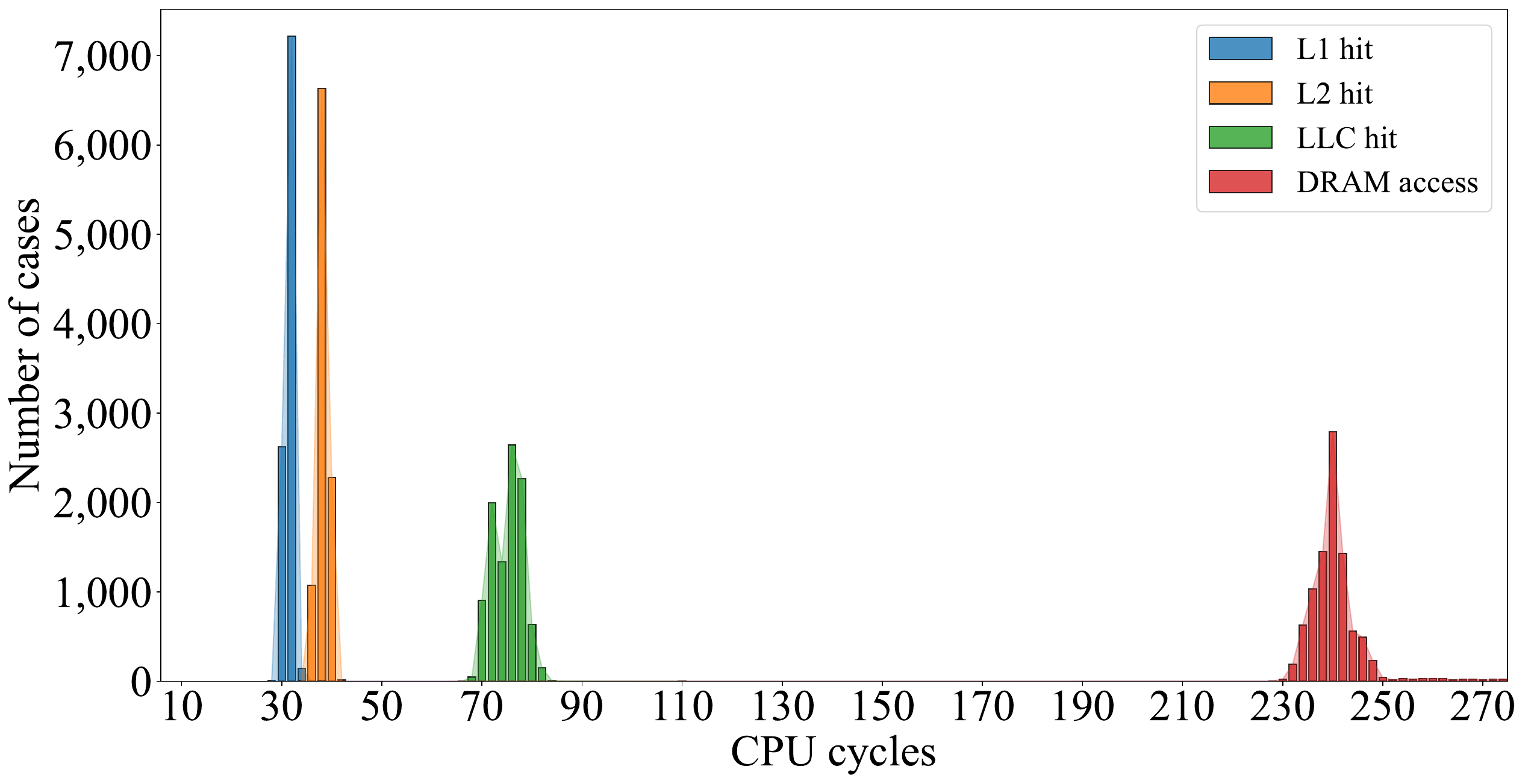}}
    \vspace{-0.2cm}
    \caption{The access latencies on target address with different types of eviction sets.}
    \label{fig:fig_a1}
    \vspace{-0.4cm}
\end{figure}
\end{document}